\documentclass[11pt]{article}

\usepackage[utf8]{inputenc}
\usepackage[OT1]{fontenc}

\usepackage{fullpage}
\usepackage{setspace}
\usepackage{natbib}
\usepackage{authblk}

\usepackage{graphicx}
\usepackage{amsmath, amsthm, amssymb}
\theoremstyle{remark}
\newtheorem{remark}{Remark}

\title{Reconstruction of Order Flows using Aggregated Data}

\author[1]{Ioane Muni Toke}
\affil[1]{MICS Laboratory, Chair Of Quantitative Finance, CentraleSupelec, France.}
\affil[1]{ERIM, University of New Caledonia, New Caledonia.}
\affil[1]{CREST, Japan Science and Technology Agency, Japan.}

\begin{document}

\maketitle

\begin{abstract}
In this work we investigate tick-by-tick data provided by the TRTH database for several stocks on three different exchanges (Paris - Euronext, London and Frankfurt - Deutsche Börse) and on a 5-year span.
We use a simple algorithm that helps the synchronization of the trades and quotes data sources, providing enhancements to the basic procedure that, depending on the time period and the exchange, are shown to be significant. We show that the analysis of the performance of this algorithm turns out to be a a forensic tool assessing the quality of the aggregated database: we are able to track through the data some significant technical changes that occurred on the studied exchanges.
We also illustrate the fact that the choices made when reconstructing order flows have consequences on the quantitative models that are calibrated afterwards on such data.
Our study also provides elements on the trade signature, and we are able to give a more refined look at the standard Lee-Ready procedure, giving new elements on the way optimal lags should be chosen when using this method. The findings are in line with both financial reasoning and the analysis of an illustrative Poisson model of the order flow.
\end{abstract}

\section{Introduction}

The development of high-frequency financial markets produces a huge amount of data every day. After the studies of monthly prices of wool in the first part of the 20th century (reported by \citet{Mandelbrot1963}), the daily indicators (open, high, low, close,...) commonly used in the 1980s and 1990s, intraday data and even (very) high-frequency data is now a common object of study for quantitative finance modelling by both academics and practitioners.
However, there is currently no large public diffusion of complete and transparent data by the exchanges, and to our knowledge there is no plan for such data sharing. Because they want to protect the identity of the market participants, exchanges are reluctant to release very detailed data. Even anonymous versions of such data might not be easily found, and, in any case, even less detailed/coarse/aggregated data is often available to academics at a non-negligible price.

The basic database available to most researchers is thus the tick-by-tick "trades and quotes" database, which reports the anonymous transactions and updates of aggregated quotes as they happen during the trading days. These are (supposedly exhaustive) snapshots of the state of the limit order book, but they do not detail the messages that trigger the transitions from one state to another. Researchers have therefore devised procedures and algorithms to infer some characteristics of these messages, by tackling for example the signature problem, which aims at determining if a given transaction has been initiated by the buying or the selling party. In this matter, the seminal work by \cite{LeeReady1991} is still a current object of study (see e.g. \cite{Easley2016} for a very recent contribution).

Recent developments in market microstructure however focus on a more detailed point of view, in which the characteristics of each order are to be studied in order to get a better grasp of the dynamics of the limit order books. Two examples of such works include limit order book modelling (see e.g. \cite{Cont2010,MuniToke2015,Huang2015} among many others) or order flow modelling with Hawkes processes (see references in Section \ref{section:SideEffects}).

This work deals with the reconstruction of a detailed order flow from a standard "trades" and "quotes" database. This exercise would be extremely simple and straightforward if the databases were perfect, i.e. if they never missed an update, if their timestamps were accurate to an infinite precision, and if all their sources were perfectly synchronized. As this is obviously not the case, the reconstruction of the order flow needs some adapted algorithms.

In Section \ref{section:Data}, we describe the database available, and in Section \ref{section:ReconstructionOrderFlow} we detail the difficulties encountered when trying to synchronize the "trades" and "quotes" sources, and the consequences on the matching procedure. We propose a matching procedure that significantly improves the basic reasoning. For the purpose of illustration, this section presents results on stocks traded in Paris in 2010 and 2011, but more extensive results are provided in Section \ref{section:EmpiricalResults}, in which the matching procedures are tested on a 5-year time window for three different exchanges (Paris - Euronext, London and Frankfurt - Deutsche Börse). It is shown that the proposed matching significantly improves the performance of the order flow reconstruction. Very interestingly, these tests reveals that technical changes that affected these exchanges during these periods leave profound marks on the tick-by-tick data, even after several steps of transformation and formatting by the exchanges and data vendors. In Section \ref{section:TradeSignature}, we use our order flow reconstruction to contribute to the assessment of the Lee-Ready algorithm of trade signature. Because we have exactly matched trades to their consequences in the order book, we know their "true" sign and can assess the performance of the Lee-Ready procedure. We show that one should not recommend an absolute lag value for the Lee-Ready procedure, as the optimal lag is time-dependent. Moreover, the study of the algorithm performance is informative on some events affecting the exchanges and even a high-level operation such as trade signature on aggregated data reveals technical changes affecting data production.
Finally, Section \ref{section:SideEffects} illustrates how the choices made when reconstructing the order flow can potentially impact the calibration of quantitative models on the reconstructed samples, underlining the importance of the proposed procedure.

\section{Description of the data}
\label{section:Data}
We use the Thomson-Reuters Tick History (TRTH) database in the form available at the Chair of Quantitative Finance in CentraleSup\'elec, Paris, France.
This database gives for a large spectrum of stocks and futures all the reported trades and all the updates of the quotes of the limit order book, up to a given number of limits (usually 5 or 10 depending on the date and stock studied).
The TRTH database identifies financial products by their (unique) Reuters Instrument Code (RIC). For the sake of illustration, 'LAGA.PA' denotes the stock Lagard\`ere traded in Paris, 'BARC.L' stands for Barclays traded in London, 'FEIZ9' is the December 2009 futures on the 3-month Euribor, etc.
The version hosted by CentraleSup\'elec spans roughly five years of data between 2008 and 2013, with several gaps.

The basic extraction\footnote{Basic extraction is done by a BNP Paribas-developed script which we do not control.} from the database gives us for any given product and trading day two files\footnote{Some files can be made available to academic researchers upon request.}. The first one, which will be called "trades" file, lists all the recorded trades (price and volume) with a timestamp in milliseconds. A "trades" file has thus three fields : "timestamp", "price" and "quantity".
The second one, which will be called "quotes" file, lists all the updates (price and volume) of the number of shares standing in the order book at the five or ten best limits around the spread. A "quotes" file has thus five fields : "timestamp", "side", "level", "price" and "quantity".

Table \ref{table:TradesFile} show the structure of the "trades" file on an actual extract of the database for the stock LAGA.PA on January 28th, 2010.
\begin{table}[htbp]
\begin{center}
\begin{tabular}{llr}
\hline
timestamp & price & quantity\\\hline
35977.101 & 27.54 & 180 \\ 
35987.244 & 27.545 & 200 \\ 
35987.244 & 27.55 & 47 \\ 
35987.244 & 27.55 & 129 \\ 
35987.247 & 27.55 & 100 \\ \hline
\end{tabular}
\end{center}
\caption{Extract from the "trades" file for the stock LAGA.PA on January 28th, 2015}
\label{table:TradesFile}
\end{table}
The "trades" file illustrated in Table \ref{table:TradesFile} is straightforward to read. First line of Table \ref{table:TradesFile} means that at time 35977.101 (in seconds, i.e. 9:59:37.101 a.m.), there has been a transaction  of 180 shares at price 27.54 EUR. In other words, each line of the "trades" file can be translated into a market order that triggers a transaction.
However, the "trades" file gives explicitly the price and the volume of a given market order, but does not say if it was a buy or a sell market order: we do not know the \emph{aggressor} in microstructure terms, i.e. the last party to enter the deal and to therefore trigger a trade. This will be addressed in Section \ref{section:TradeSignature}.

Table \ref{table:QuotesFile} show the structure of the "quotes" file on an actual extract of the database for the stock LAGA.PA on January 28th, 2010.
\begin{table}[htbp]
\begin{center}
\begin{tabular}{lcclr}
\hline
timestamp & side & level & price & quantity \\ \hline
36003.97 & A & 3 & 27.585 & 697 \\ 
36003.97 & A & 3 & 27.585 & 177 \\ 
36004.067 & A & 3 & 27.59 & 391 \\ 
36004.067 & A & 4 & 27.595 & 311 \\ 
36004.067 & A & 5 & 27.6 & 427 \\ 
36004.067 & A & 6 & 27.605 & 1688 \\ 
36004.067 & A & 7 & 27.61 & 586 \\ 
36004.067 & A & 8 & 27.615 & 677 \\ 
36004.067 & A & 9 & 27.62 & 1999 \\ 
36004.067 & A & 10 & 27.625 & 568 \\ 
36004.613 & A & 6 & 27.605 & 2315 \\ 
36005.314 & B & 1 & 27.55 & 6829 \\ \hline
\end{tabular}
\end{center}
\caption{Extract from the "quotes" file for the stock LAGA.PA on January 28th, 2015}
\label{table:QuotesFile}
\end{table}
The penultimate line of Table \ref{table:QuotesFile} means, by comparing it to the sixth line from the top, that there has been a increase of the offered volume (new limit order) at time 10:00:04.613 at price 27.605 EUR of size 2315-1688=627.
The second line of Table \ref{table:QuotesFile} means, by comparing to the first line of the extract, that there has been a decrease of the offered volume (cancellation of a standing limit order) at time 10:00:03.97 at price 27.585 EUR of size 697-177=520.

In short, our data is tick-by-tick data in the sense that all event modifying the order book are supposed to be individually reported, and aggregated in the sense that we do not have access to the granularity of the messages/orders but to the aggregated volume of the sizes of the orders standing in the limit order book.

\section{Reconstruction of the order flow}
\label{section:ReconstructionOrderFlow}

Our goal is to reconstruct the precise \emph{order flow}, i.e. the complete sequence of market orders, limit orders and cancellations along with their price and volume that lead to the generation of given "trades" and "quotes" files. The general algorithm for the determination of the order flow would be straightforward if the database were perfectly coherent, but inconsistencies exist, mainly because of the non-synchronization of the timestamps of the trades and quotes files, and maybe because of probable inaccuracies in the data building process.

\subsection{Sample description}
For the purpose of illustration while presenting the construction of the algorithm, we randomly select on our database seven liquid stocks from the CAC 40 (Air Liquide AIRP.PA, Alstom ALSO.PA, BNP Paribas BNPP.PA, Bouygues BOUY.PA, Carrefour CARR.PA, Electricité de France EDF.PA, Lagardère LAGA.PA). For each stock we extract from the database the "quotes" and "trades" files as described above from January 17th, 2011.
Please note that this sample is selected for illustration only, and that Section \ref{section:EmpiricalResults} will provide more detailed results, for more stocks and an extensive range of dates.

\subsection{Parsing the "quotes" file}
We start with an empty structure of $2N$ pairs "price"-"quantity" representing the state of the order book, i.e. the number of share available at the $N$ best bid and ask, and their price. ($N=5$ or $10$ limits are usually available, depending on the RIC and trading day).
We scan line by line the "quotes" file. The reading of the first lines allow us to initialize the state of the order book. Each following line is an update of a given volume and price: in a first approach, each line is thus a new limit order (increase of the number of shares available at a given price), or a cancellation of an existing limit order (decrease of the number of shares available at a given price), as in the two examples provided with Table \ref{table:QuotesFile}.

When parsing the "quotes" file, one has to be cautious of "shifts". For example, if a limit order is submitted within the spread or at a price within the current $N$ best quotes where no previous limit order were standing, then it creates a new level, and all subsequent levels are re-numbered. This gives multiple updates messages in the "quotes" file that are not actual orders. An example is provided in Table \ref{table:QuotesShiftComplete}. The first ten lines describe the current state of the order book. The eleventh line shows that a new ask limit order with size 66 is submitted inside the spread at price 27.52 EUR. Then the nine following lines do \emph{not} describe any limit order, but simply update the level numbering (the new level 2 is the level 1, and so on). When parsing these lines, the available volume at level $n$ is to be compared to the previous volume at level $n-1$, in order to track potential changes.
\begin{table}[htbp]
\begin{center}
\begin{tabular}{lcclr}
\hline
timestamp & side & level & price & quantity \\ \hline
 34819.37 & A & 1 & 27.54 & 326 \\ 
 34819.37 & A & 2 & 27.545 & 530 \\ 
 34819.37 & A & 3 & 27.55 & 989 \\ 
 34819.37 & A & 4 & 27.555 & 318 \\ 
 34819.37 & A & 5 & 27.56 & 79 \\ 
 34819.37 & A & 6 & 27.565 & 275 \\ 
 34819.37 & A & 7 & 27.57 & 468 \\ 
 34819.37 & A & 8 & 27.58 & 100 \\ 
 34819.37 & A & 9 & 27.585 & 612 \\ 
 34819.37 & A & 10 & 27.59 & 1638 \\ 
 34819.37 & A & 1 & 27.52 & 66 \\ 
 34819.37 & A & 2 & 27.54 & 326 \\ 
 34819.37 & A & 3 & 27.545 & 530 \\ 
 34819.37 & A & 4 & 27.55 & 989 \\ 
 34819.37 & A & 5 & 27.555 & 318 \\ 
 34819.37 & A & 6 & 27.56 & 79 \\ 
 34819.37 & A & 7 & 27.565 & 275 \\ 
 34819.37 & A & 8 & 27.57 & 468 \\ 
 34819.37 & A & 9 & 27.58 & 100 \\ 
 34819.37 & A & 10 & 27.585 & 612 \\ \hline
\end{tabular}
\end{center}
\caption{"Shift" of quotes due to a limit submitted at a price where no previous limit order was standing, inside the spread in this example. Extract from the "quotes" file for LAGA.PA on January 28th, 2010.}
\label{table:QuotesShiftComplete}
\end{table}

Since we observe a fixed finite number of prices $N$ (usually equal to $10$), shifts may make some information disappear: in table \ref{table:QuotesShiftComplete} for example, the fact that 1638 shares are available at price 27.59 on the ask side is lost at the update because of a left shift of this side of the book. By a reciprocal mechanism, a right shift of the ask side (or a left shift of the bid side) may bring new information to light, that is \emph{not} to be interpreted as a new order. In table \ref{table:PerfectTradeMatching} below for example, the last line stating that 598 shares are available at 27.095 on the tenth limit of the bid is not a new limit order --- despite being a new information --- because this price was not monitored before.

When the whole "quotes" file is parsed, we have a sequence of limit orders and cancellations --- no market orders yet --- with their timestamp to the millisecond, side, level, price and volume.

\subsection{Parsing the "trades" file in the perfect case}

The second step of the order flow reconstruction deals with the parsing of the "trades" file. If everything were perfectly recorded, then each line of the "trades" file should represent a market order, and at the exact same timestamp, we should see in the "quotes" file a decrease of the quantity at the same price and with the same volume. 
In other words, each line of the "trades" file should match a cancel order of the same volume, at the same price and timestamp in the order flow resulting from the parsing of the "quotes" file.
An example of this perfect case is documented in Table \ref{table:PerfectTradeMatching}.
\begin{table}[htbp]
\begin{center}
\begin{tabular}{llr}
\hline
timestamp & price & quantity\\\hline
32472.252 & 27.32 & 267 \\ \hline
\end{tabular}

\begin{tabular}{lcclr}
\hline
timestamp & side & level & price & quantity \\ \hline
32472.086 & B & 1 & 27.32 & 267 \\ 
32472.086 & B & 2 & 27.31 & 500 \\ 
32472.086 & B & 3 & 27.29 & 585 \\ 
32472.086 & B & 4 & 27.285 & 127 \\ 
32472.086 & B & 5 & 27.27 & 500 \\ 
32472.086 & B & 6 & 27.2 & 300 \\ 
32472.086 & B & 7 & 27.16 & 500 \\ 
32472.086 & B & 8 & 27.155 & 200 \\ 
32472.086 & B & 9 & 27.15 & 1750 \\ 
32472.086 & B & 10 & 27.1 & 223 \\ 
32472.086 & B & 2 & 27.31 & 710 \\ 
32472.252 & B & 1 & 27.31 & 710 \\ 
32472.252 & B & 2 & 27.29 & 585 \\ 
32472.252 & B & 3 & 27.285 & 127 \\ 
32472.252 & B & 4 & 27.27 & 500 \\ 
32472.252 & B & 5 & 27.2 & 300 \\ 
32472.252 & B & 6 & 27.16 & 500 \\ 
32472.252 & B & 7 & 27.155 & 200 \\ 
32472.252 & B & 8 & 27.15 & 1750 \\ 
32472.252 & B & 9 & 27.1 & 223 \\ 
32472.252 & B & 10 & 27.095 & 598 \\ \hline
\end{tabular}
\end{center}
\caption{Perfect match between the "trades" and quotes" files. The line of the "trades" extract (top table) is perfectly matched with the same timestamp in the extract of the "quotes" file (bottom table). Stock LAGA.PA on January 28th, 2010.}
\label{table:PerfectTradeMatching}
\end{table}
After parsing the extract of the "quotes" file in Table \ref{table:PerfectTradeMatching}, we would have the following order flow :
\begin{center}
\begin{tabular}{lccclr}
\hline
32472.086 &  LIMIT & B & 2 & 27.31 & 210 \\
32472.252 &  CANCEL & B & 1 & 27.32 & 267 \\ \hline
\end{tabular}
\end{center}
But after parsing the "trades" file, we conclude that the update at time 32472.252 is not a cancellation, but the result of a market order. We thus change the order flow to :
\begin{center}
\begin{tabular}{lccclr}
\hline
32472.086 &  LIMIT & B & 2 & 27.31 & 210 \\
32472.252 &  MARKET & B & 1 & 27.32 & 267 \\ \hline
\end{tabular}
\end{center}

\subsection{Basic approach of trade matching in the general case}

The perfect case described above is however extremely rarely observed. On our 7-stock 1-day illustrative sample, the perfect case occurs at a frequency lower than 0.001 (but one should refrain from generalizing this figure, as Section \ref{section:EmpiricalResults} will show that matching results are strongly exchange and time dependent). 
The cause of this situation is that the clock used to timestamp the "trades" and "quotes" files are not synchronised. Therefore, it is very rare to exactly match a corresponding market order read in the "trades" file to a cancellation read in the "quotes" file. Most of the time, the transaction is recorded in the "trades" file and the updates are published in the "quotes" file with a few milliseconds lag.

Therefore in a first approach of trade matching, when we parse the "trades" file and look for a matching cancellation in our order flow constructed from the "quotes" file, with the exact same price and volume, we look not only at the exact transaction timestamp, but within $-\delta$ and $+\delta$ seconds around this transaction timestamp. In the event of multiple matches possible, we choose the one which minimizes the timestamp difference. This procedure will called "Matching 1" from now on.

\begin{figure}
\begin{center}
\includegraphics[scale=0.5, page=1]{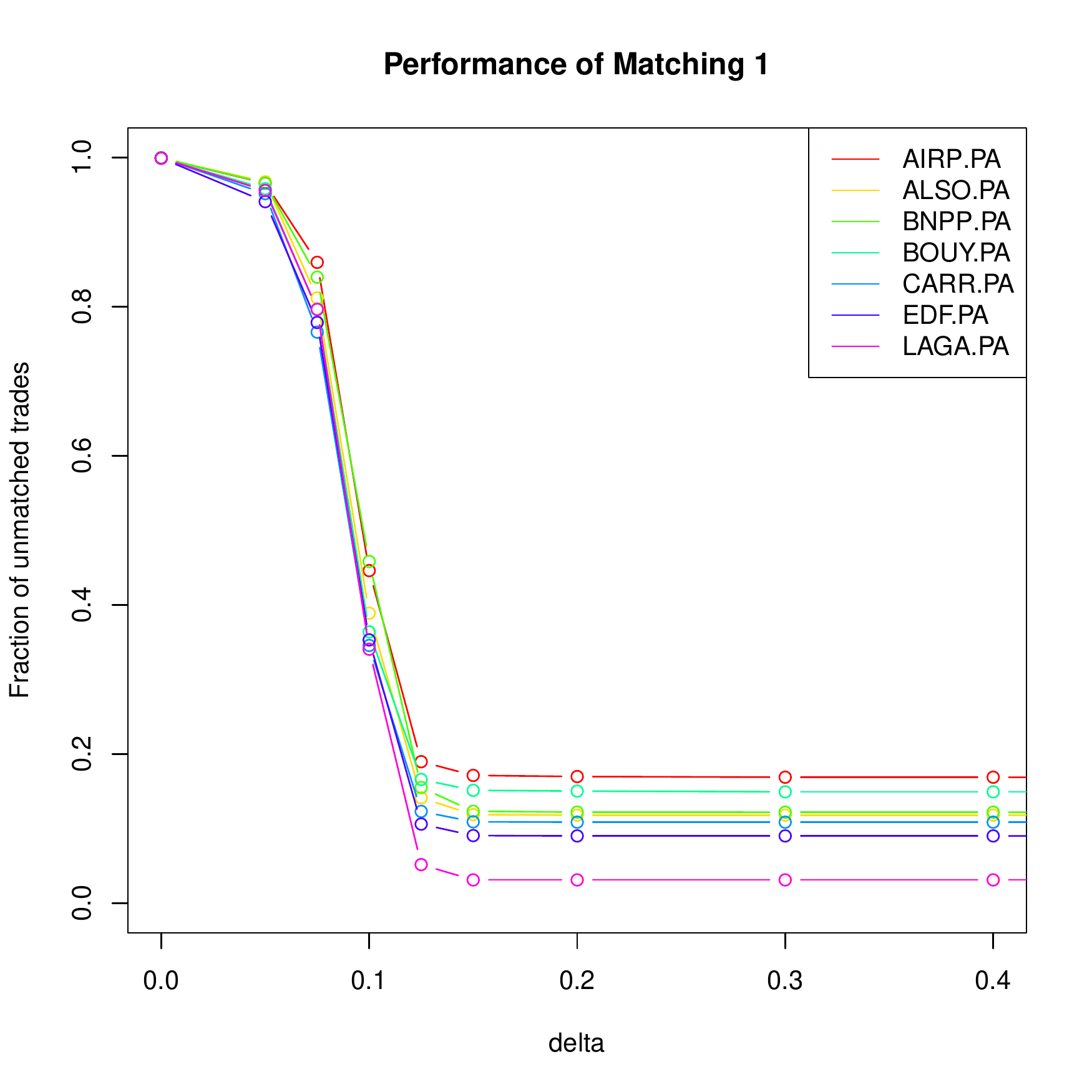}
\end{center}
\caption{Proportion of trades left unmatched by Matching 1. All trades reported in the database on Jan. 17th, 2011 between 09:05 and 17:20 are considered for matching.}
\label{figure:Matching1-Performance}
\end{figure}

Figure \ref{figure:Matching1-Performance} plots the proportion of trade left unmatched after the Matching 1 procedure has been applied to the sample.
It appears that the performance of the matching is similar for all stocks. As expected, when $\delta$ increases, the matching performance improves, but unexpectedly it reaches a plateau for all stocks for $\delta\approx150-200$ ms. Beyond this value, increasing $\delta$ has no effect of the performance, and the Matching 1 procedure leaves unmatched between roughly $5-20\%$ of the trades reported in the trades file, in this illustrative sample.

\begin{figure}
\begin{center}
\includegraphics[scale=0.5, page=1]{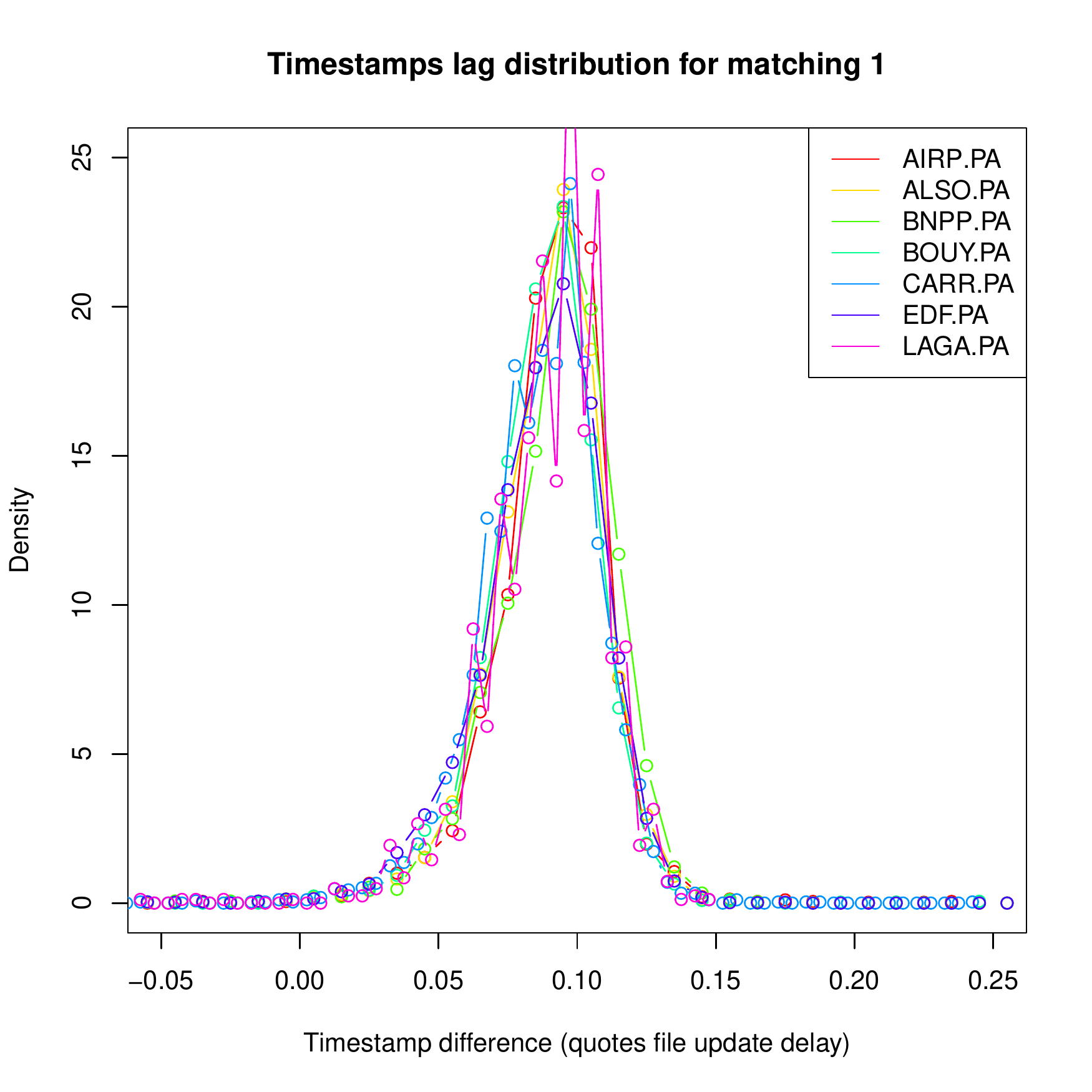}
\end{center}
\caption{Empirical density of the time lag $\Delta\tau$ between a transaction recorded in the "trades" file and its matching cancellation in the "quotes" file with the "Matching 1" procedure with $\delta=0.4$.}
\label{figure:TimeLagMatching1}
\end{figure}

For a more precise view on this phenomenon, we plot in Figure \ref{figure:TimeLagMatching1} the empirical distribution of the measured time lags $\Delta\tau=\tau_q-\tau_t$ where $\tau_t$ is the time stamp of a trade in the "trades" file, and $\tau_q$ the time stamp of its matched modification in the "quotes" file.
This distribution has a mainly positive support, indicating that the updates of the "quotes" files are most of the time following the report of the transaction in the "trades" file. Furthermore, all the empirical densities drop to $0$ around $150$ ms, indicating that it is rare to match an order more that 150 ms later in the "quotes" file.

\subsection{Enhanced approach of trade matching}

The previous matching procedure leaves quite a large number of unmatched transactions. It is indeed quite frequent that consecutive lines in the "trades" file lead to only one update in the "quotes" files.
The common explanation for this phenomenon is that the exchange issues one message for each pending limit order that is (partially) matched by an incoming market order.
In this case, one has to aggregate consecutive lines of the "trades" file at the same price to find the corresponding cancellation in the "quotes" file.
This is illustrated in Table \ref{table:TradesAggregation}.
\begin{table}[htbp]
\begin{center}
\begin{tabular}{llr}
\hline
timestamp & price & quantity\\\hline
32951.412 & 27.45 & 300 \\ 
32951.412 & 27.45 & 182 \\ \hline
\end{tabular}

\begin{tabular}{lcclr}
\hline
timestamp & side & level & price & quantity \\ \hline
32951.419 & A & 1 & 27.45 & 482 \\
32951.419 & A & 2 & 27.455 & 730 \\ 
32951.419 & A & 3 & 27.465 & 200 \\ 
32951.419 & A & 4 & 27.47 & 200 \\ 
32951.419 & A & 5 & 27.475 & 200 \\ 
32951.419 & A & 6 & 27.48 & 279 \\ 
32951.419 & A & 7 & 27.485 & 1813 \\ 
32951.419 & A & 8 & 27.495 & 529 \\ 
32951.419 & A & 9 & 27.5 & 3030 \\ 
32951.419 & A & 10 & 27.505 & 200 \\ 
32951.419 & A & 1 & 27.455 & 730 \\ \hline
\end{tabular}
\end{center}
\caption{Aggregation of two lines of the "trades" file (top table) to match the updates of the "quotes" file (bottom table). Stock LAGA.PA on January 28th, 2010}
\label{table:TradesAggregation}
\end{table}
The result of the parsing of the extract of the "quotes" file in Table \ref{table:TradesAggregation} is the cancel order (CANCEL, 32951.419, A, 1, 27.45, 482). After parsing the "quotes" file, we conclude that it is in fact the recording with a 7 milliseconds delay of a market order that was recorded in the "trades" file as two transactions.

The difficulty is that there is no simple rule to make this procedure automatic, since all market orders at the same timestamp do not necessarily have to be aggregated to match the "quotes" file. In order to tackle these difficulties, we enhance the matching procedure: when parsing the "trades" file, we group consecutive lines (transactions) with the same price and the same timestamp (for now ; see below) into a batch. Then for each constructed batch, we test the matching of all its possible divisions into partitions of consecutive lines, and keep as our solution the matching that matches the highest number of lines of the original "trades" file. For example, when parsing a "trades" file, if we find three transactions at the same timestamp, as follows (synthetic example, not real data):
\begin{center}
\begin{tabular}{llr}
\hline
timestamp & price & quantity\\\hline
36000.000 & 20.00 & 100 \\ \hline
36000.000 & 20.00 & 50 \\ \hline
36000.000 & 20.00 & 200 \\ \hline
\end{tabular}
\end{center}
then in addition to the matching of these three lines consecutively, we will also test the matching the following of the arrangements of the transactions :
\begin{center}
\begin{tabular}{llr}
\hline
timestamp & price & quantity\\\hline
36000.000 & 20.00 & 150 \\ \hline
36000.000 & 20.00 & 200 \\ \hline
\end{tabular}
\end{center}
\begin{center}
\begin{tabular}{llr}
\hline
timestamp & price & quantity\\\hline
36000.000 & 20.00 & 100 \\ \hline
36000.000 & 20.00 & 250 \\ \hline
\end{tabular}
\end{center}
\begin{center}
\begin{tabular}{llr}
\hline
timestamp & price & quantity\\\hline
36000.000 & 20.00 & 350 \\ \hline
\end{tabular}
\end{center}

\begin{figure}
\begin{center}
\includegraphics[scale=0.5, page=2]{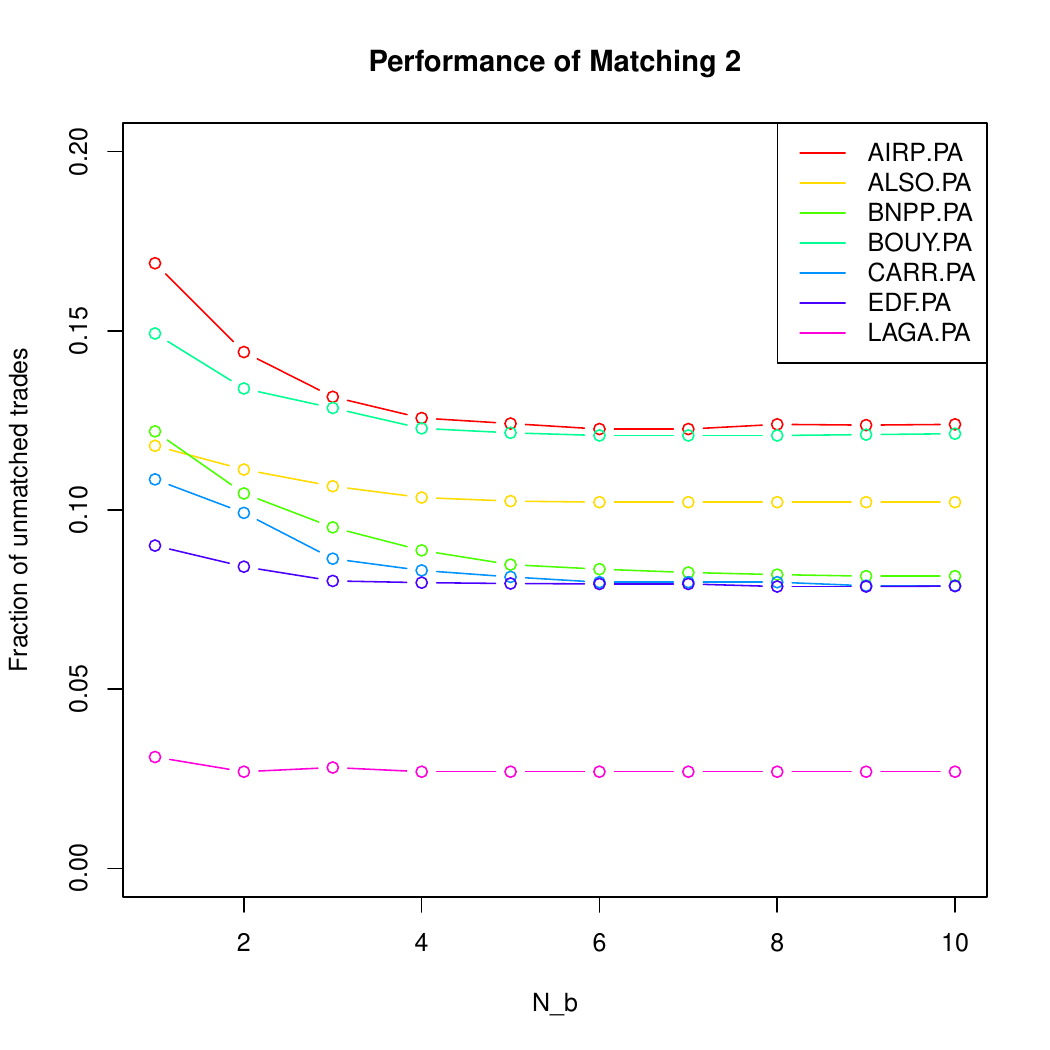}
\end{center}
\caption{Proportion of trades left unmatched by Matching 2. All trades reported in the database on Jan. 17th, 2011 between 09:05 and 17:20 are considered for matching. In this example, $\delta=0.4$.}
\label{figure:Matching2-Performance}
\end{figure}

To illustrate the performance increase due to this enhancement, Figure \ref{figure:Matching2-Performance} plots the fraction of trades left unmatched by this Matching 2 procedure as a function of $N_b$, the maximum size allowed for a single batch.
When $N_b=1$ (first point on the left), the matching is identical to the Matching 1 procedure ($\delta=0.4$ in the example plotted). As expected, when $N_b$ increases, the matching is improved : on our sample, a minimum of $12\%$ (LAGA.PA) and up to $33\%$ (BNPP.PA) of previously unmatched trades are now matched to the limit order book, which for BNPP.PA or AIRP.PA represents more than $4\%$ of the total transactions. Note however that on this sample increasing the size of the batch above 5 or 6 lines does not have any significant impact on the performance.

Among the orders that are still left unmatched by Matching 2, another difficulty is identified by a careful observation of data files : market orders that do not have exactly the same timestamp sometimes have to be aggregated to match the "quotes" file.
This is illustrated in Table \ref{table:LooseTradesAggregation}.
\begin{table}[htbp]
\begin{center}
\begin{tabular}{llr}
\hline
timestamp & price & quantity\\\hline
33095.296 & 27.51 & 100 \\ 
33095.296 & 27.51 & 202 \\ 
33095.296 & 27.51 & 303 \\ 
33095.296 & 27.51 & 486 \\ 
33095.296 & 27.51 & 334 \\ 
33095.296 & 27.535 & 210 \\ 
33095.299 & 27.535 & 140 \\ \hline
\end{tabular}
\\
\begin{tabular}{lcclr}
\hline
timestamp & side & level & price & quantity \\ \hline
33046.84 & A & 1 & 27.535 & 3951 \\
\multicolumn{5}{c}{[\ldots]}\\
33085.836 & B & 1 & 27.51 & 2264 \\ 
33095.297 & B & 1 & 27.51 & 839 \\ 
33095.31 & A & 1 & 27.535 & 3601 \\ \hline
\end{tabular}
\end{center}
\caption{Aggregation of lines of the "trades" file with different time stamps (top table) to match the updates of the "quotes" file (bottom table). Stock LAGA.PA on January 28th, 2010.}
\label{table:LooseTradesAggregation}
\end{table}
In Table \ref{table:LooseTradesAggregation}, the five first line of the extract of the "trades" file have to be aggregated to match the order (33095.297, CANCEL, B, 1, 27.51, 1425), with a 1 millisecond time lag. But the last two orders, although with two different timestamps separated with 3 milliseconds have to be aggregated to match the order (CANCEL, A, A, 27.535, 350), reported with a 11 milliseconds delay from one order, and 14 milliseconds delay from the other.

\begin{figure}
\begin{center}
\includegraphics[scale=0.5, page=3]{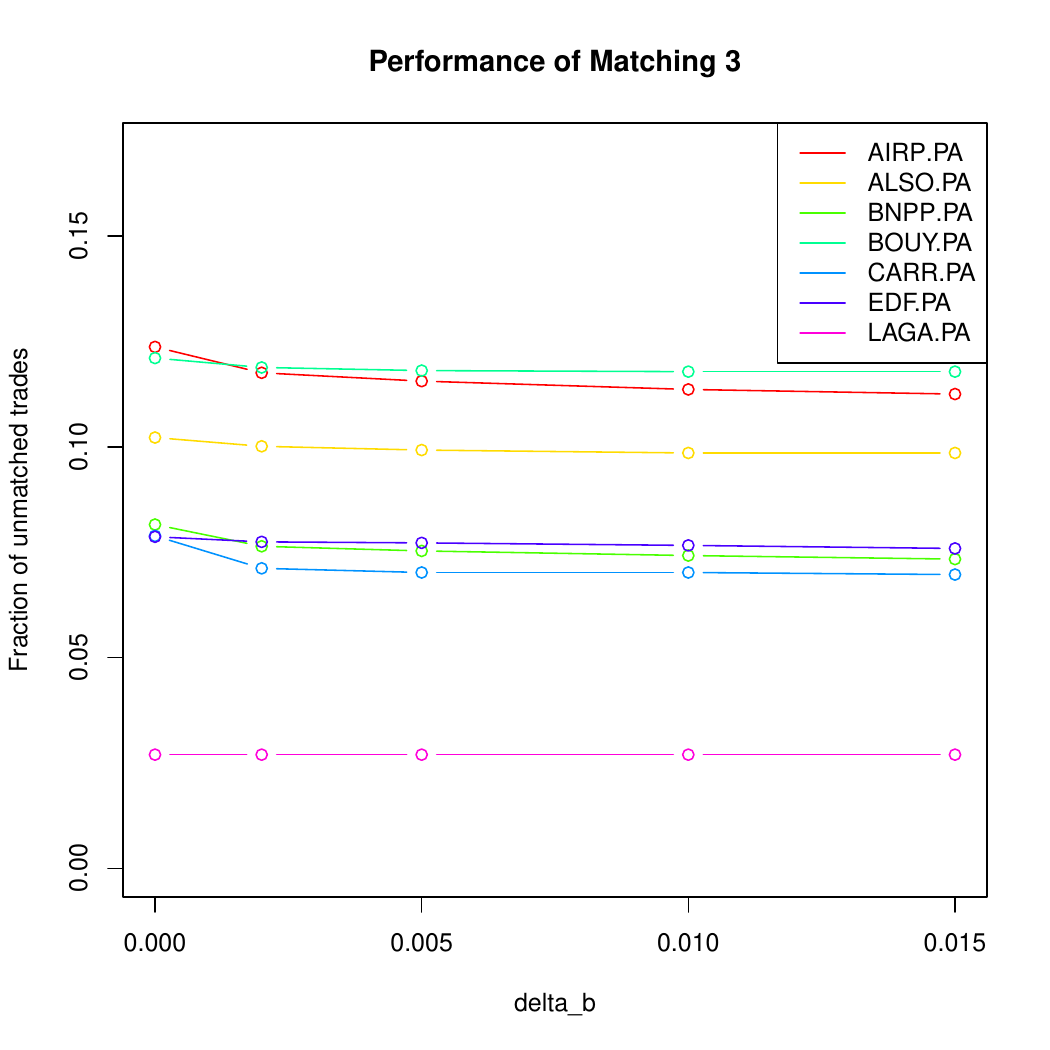}
\end{center}
\caption{Proportion of trades left unmatched by Matching 3. All trades reported in the database on Jan. 17th, 2011 between 09:05 and 17:20 are considered for matching. In this example, $\delta=0.4, N_b=9$.}
\label{figure:Matching3-Performance}
\end{figure}

In order to track these special cases, we enhance the Matching 2 procedure described above by grouping into the same batch lines that have a time stamp \emph{close} to the first time stamp read, but not necessarily equal. The parameter measuring the maximum allowed time lag of a single batch will be denoted $\delta_b$.
To illustrate the performance increase due to this enhancement, Figure \ref{figure:Matching3-Performance} plots the fraction of trades left unmatched by this Matching 3 procedure as a function of $\delta_b$.
In this sample it appears that the transactions that are to be aggregated have a very close time stamp, and that the typical time lag to be considered  is 2-3 ms. For higher values of $\delta_b$ no matching improvements is observed.
The improvement of Matching 3 allows to match between $3\%$ (ALSO.PA, BOUY.PA, EDF.PA) to more that $10\%$ of the previously unmatched trades (CARR.PA, BNPP.PA), except for LAGA.PA whose performance is not improved. However, this improvements represent in any case less than $1\%$ of the total transactions observed.

Table \ref{table:MatchingPerformanceSummary} summarizes the performances of the matching procedures described  above.
\begin{table}[htbp]
\begin{center}
\begin{tabular}{|l|r|r|r|}
\hline
 & \multicolumn{1}{l|}{Matching 1} & \multicolumn{1}{l|}{Matching 2} & \multicolumn{1}{l|}{Matching 3} \\ \hline
AIRP.PA & 83.11\% & 87.63\% & 88.75\% \\ \hline
ALSO.PA & 88.21\% & 89.78\% & 90.15\% \\ \hline
BNPP.PA & 87.81\% & 91.85\% & 92.67\% \\ \hline
BOUY.PA & 85.07\% & 87.90\% & 88.22\% \\ \hline
CARR.PA & 89.15\% & 92.11\% & 93.03\% \\ \hline
EDF.PA & 90.99\% & 92.14\% & 92.41\% \\ \hline
LAGA.PA & 96.89\% & 97.30\% & 97.30\% \\ \hline
\end{tabular}
\caption{Percentages of transactions for the "trades" file matched by the different matching procedures described in the text}
\label{table:MatchingPerformanceSummary}
\end{center}
\end{table}
In short, on this 7-stock sample, our detailed algorithm allows in average the matching of roughly 92\% of the trades reported in the "trades" file, which is an average improvement of $3\%$ of the total transactions from the standard Matching 1 procedure.
This may not sound much but remember that this small 1-day sample has been used just for the purposes of illustration of the matching procedure. Much detailed results are provided in the next sections, showing that for some stocks and date the enhanced matching (Matching 3) is crucial.

\subsection{Sources of unmatched trades}

Several other difficulties might be treated by further enhancing the above matching procedure, but at probable high cost with marginal enhancement.
A manual observation of the files show that some of the orders that remain unmatched truly appear "out of bound", with quantities or prices that are not at all observed in the limit order book. For many unmatched orders however, only a small volume discrepancy prevents a matching: even if the price is equal to the best bid and we observe cancellations at the best bid, we cannot exactly equate their volumes, no matter how we aggregate the information. In this case, no exact matching can be obtained.

We can only speculate about the sources of these roughly $10\%$ of trades in average that are not matched. The most obvious explanation would be that the reported transactions are executed against hidden liquidity, which is allowed by many exchanges and on Euronext Paris takes the form of "iceberg" orders. Only a fraction of the size of an iceberg order is displayed on the order book (for example assume $100$ shares are displayed alone at the best quote out of a full size of $1000$). When a market order occurs and eats part of this liquidity (let us say $50$ shares), the quotes are updated as usual (and we observe a decrease of $50$ shares, which is easily matched). But when a market order eats more than the displayed quantity (let us say $125$ shares), then it is executed against the first $100$ shares, which are immediately replaced by $100$ previously hidden shares, and the last $25$ shares are executed against the newly displayed quantity. If there is one update for this single market order, then the final observation is an decrease of $25$ shares. That would explain the fact precise quantities fail to match despite the observations of trades and corresponding cancellations at the same price.
Another explanation would be that some lines report transactions that occurred outside of the limit order book and therefore they have no effect on the "quotes" file. The inaccuracies here would be in the "trades" files, since such trades are supposed to be flagged as "off the book". 

\begin{figure}
\begin{center}
\begin{tabular}{cc}
\includegraphics[scale=0.43, page=4]{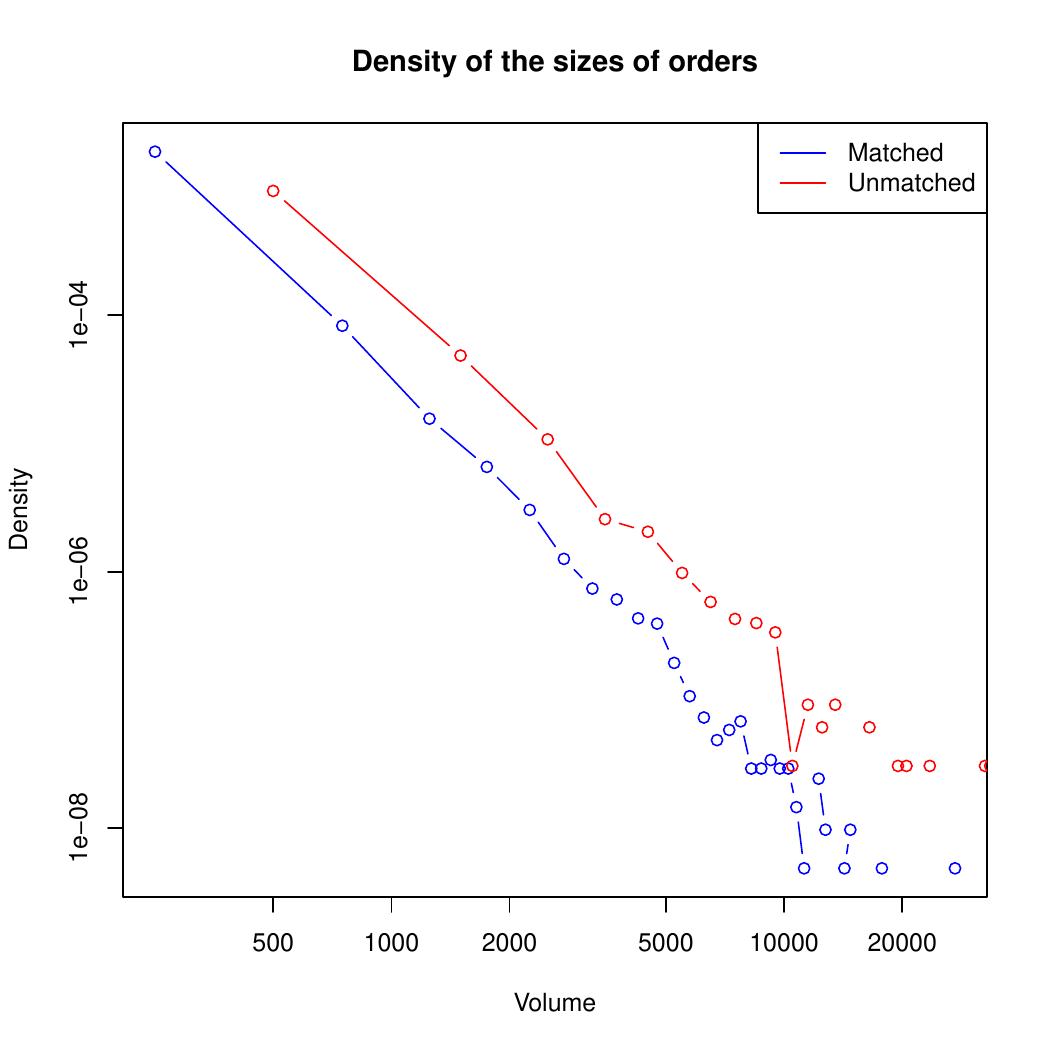}
& 
\includegraphics[scale=0.43, page=7]{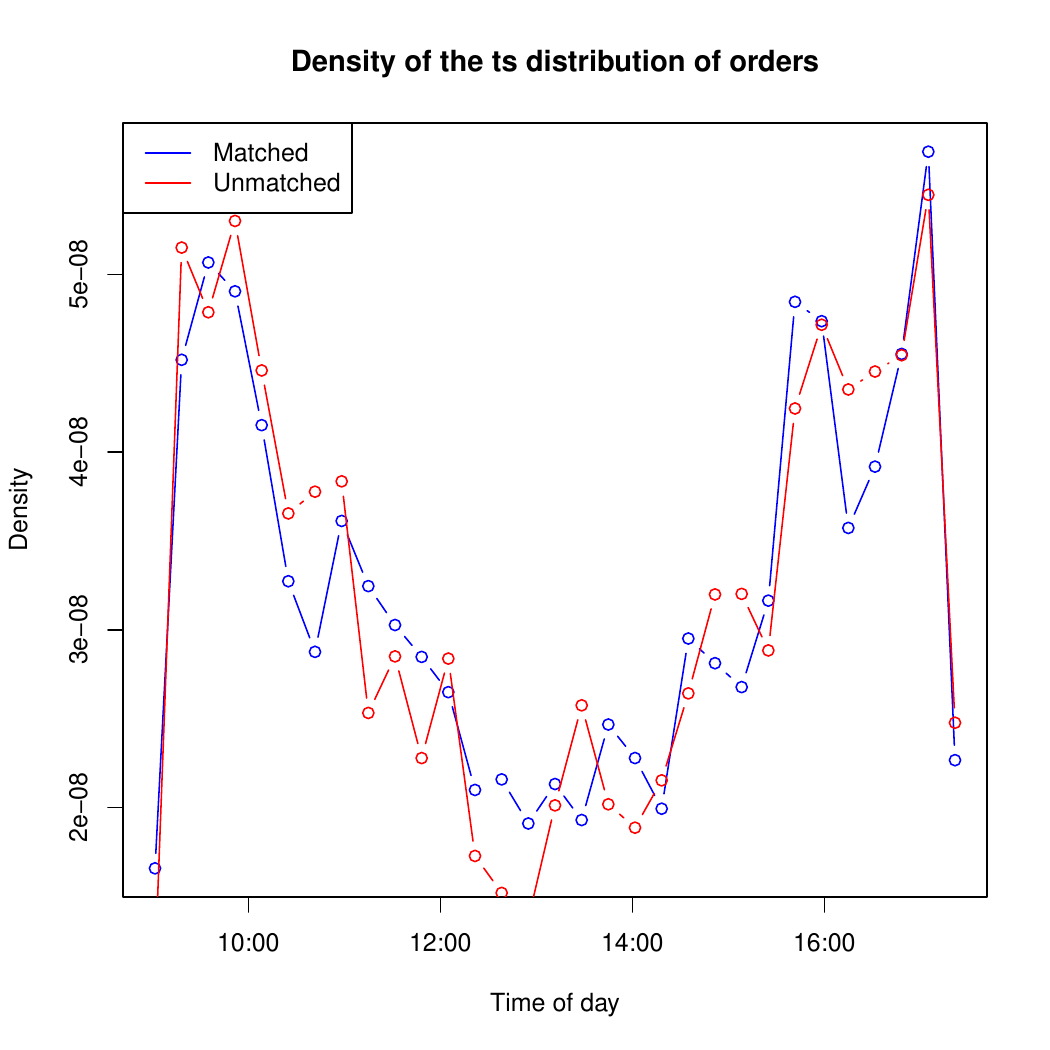}
\end{tabular}
\end{center}
\caption{Empirical density distributions of the sizes (left, in log-log scale) and the timestamps (right) of the unmatched trades, compared to the matched ones. In this example, Matching 3 is used for the stock BNPP.PA on the whole month of January 2011.}
\label{figure:UnmatchedTrades}
\end{figure}

For the sake of completeness, Figure \ref{figure:UnmatchedTrades} plots the empirical density distributions of the sizes and the timestamps of the unmatched trades, which are the only two information we have on these transactions.
On these figures we observe that unmatched trades occur all day, and that the density of their timestamp is similar to the one of the unmatched trades, following a well-known U-shaped pattern indicating a daily seasonality. Therefore we cannot link unmatched trades to specific periods of the day.

However, we observe that the sizes of the unmatched orders are significantly larger than the sizes of matched orders (a Kolmogorov-Smirnov test confirms this observation).
This may be in line with both previous assumptions. As for hidden liquidity, large market orders are more likely to fully eat the displayed chunk of an iceberg order, and therefore trigger liquidity that prevents the exact size matching. This is also coherent with "off the book transactions", since one reason to trade outside the book might be to trade larger volumes. 
Finally, this finding about size might suggest that these orders could eventually be matched by the aggregation of several quotes modifications. A manual parsing of the files does not reveal systematic examples of this behaviour, but this eventuality should not be completely ruled out yet.

\section{Empirical results for order flow reconstruction}
\label{section:EmpiricalResults}

In this section we present extensive results of our matching algorithm for multiple stocks and trading days.

\subsection{Evolution of the performance through time -- Data quality}
\label{subsection:MatchingPerformanceThoughTime}

In this section we test our algorithms on several European exchanges (Euronext - Paris Stock Exchange, London Stock Exchange, Deutsche Börse - Frankfurt Stock Exchange) on the whole range available in our database, from January 2008 to June 2013, with a large gap between June 2008 and March 2009.

\begin{figure}
\begin{center}
\begin{tabular}{cc}
\includegraphics[scale=0.43, page=1]{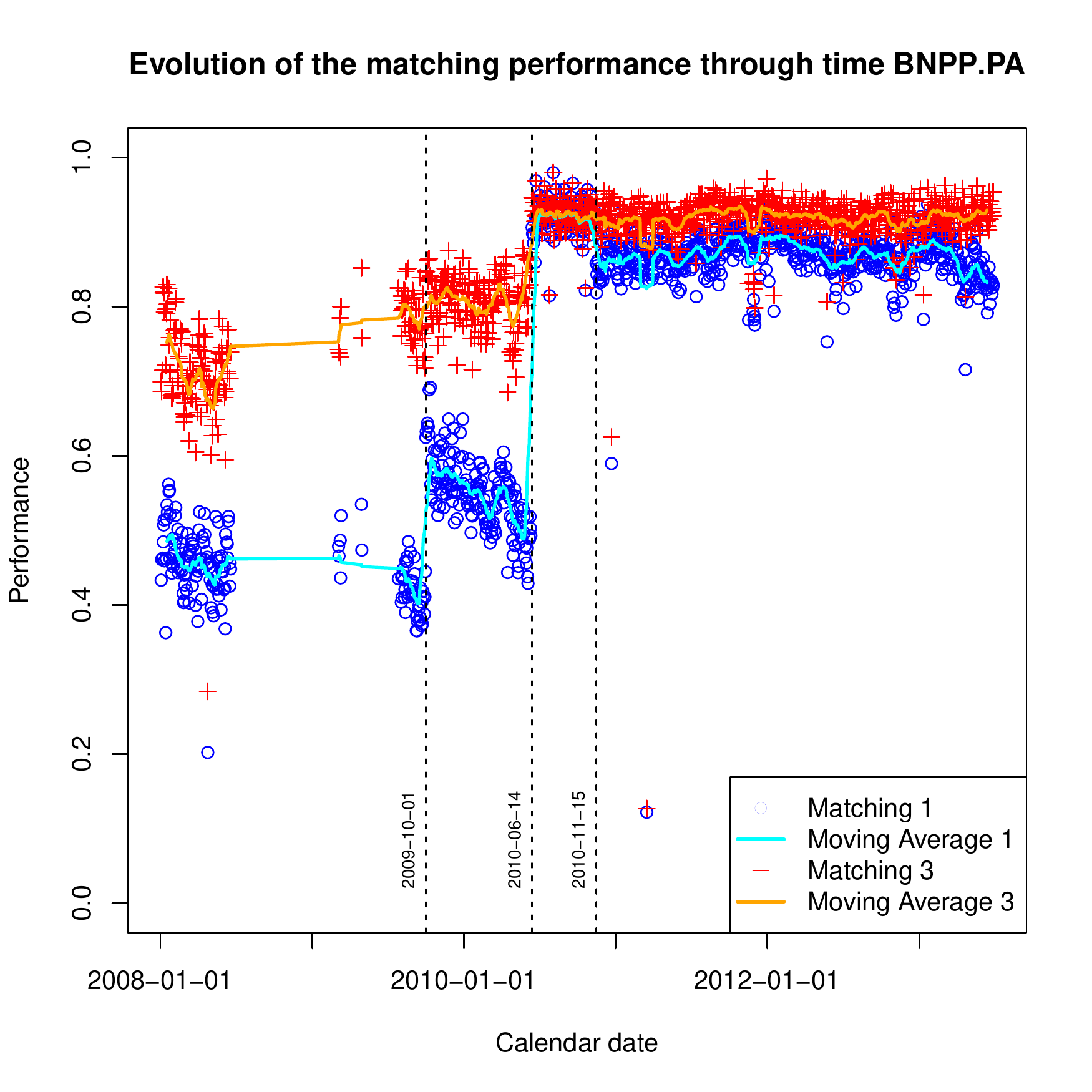}
&
\includegraphics[scale=0.43, page=1]{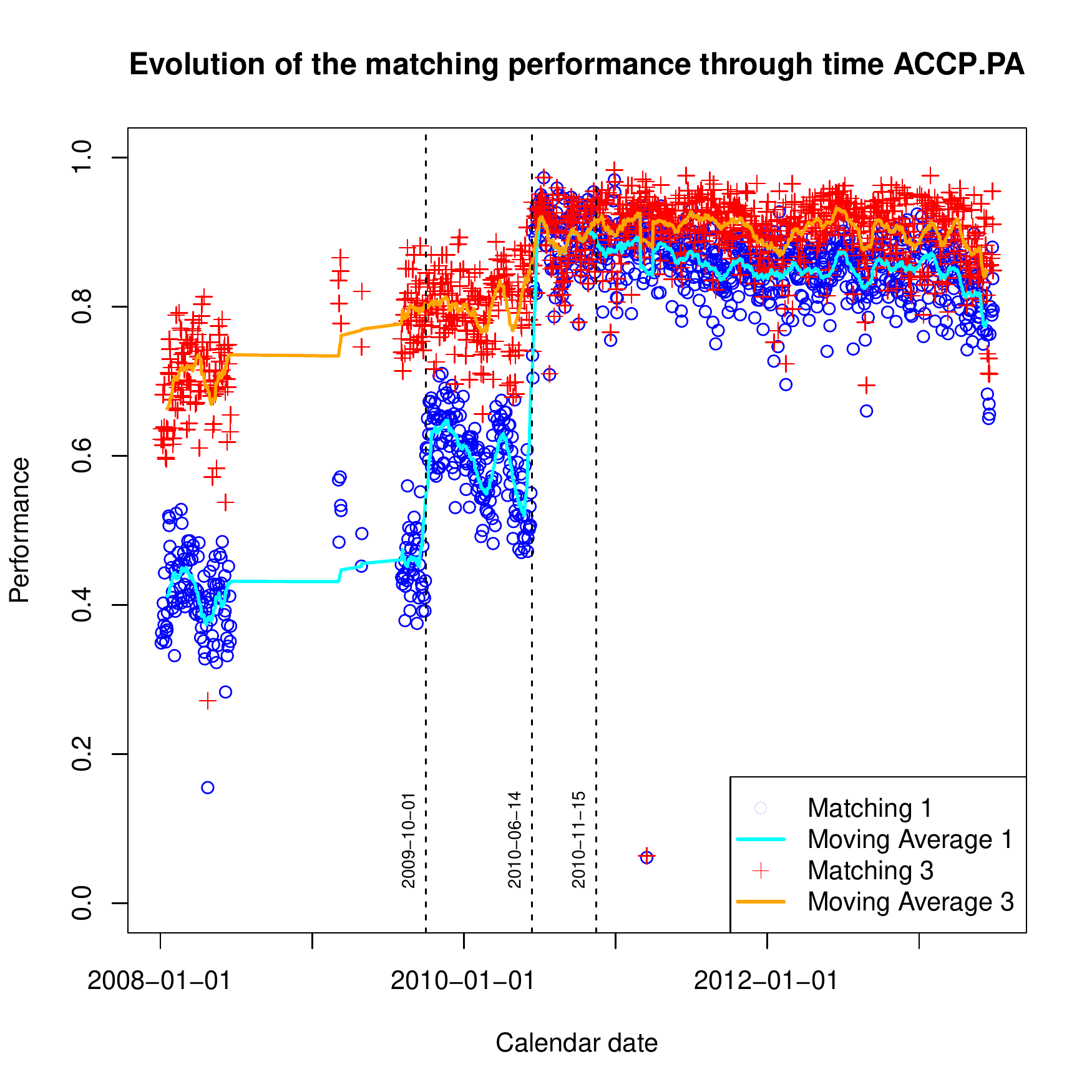}
\end{tabular}
\end{center}
\caption{Performance of the basic and enhanced matching algorithms for the stock BNPP.PA (left) and ACCP.PA (right) from Jan, 1st, 2008 to June, 30th, 2013, with gaps.}
\label{figure:PerformanceThoughTime}
\end{figure}

Results for the stocks BNP.PA (BNP Paribas) and ACCP.PA (Accor Group), both components of the CAC 40, are presented for the basic matching (Matching 1, $\delta=0.4$) and the enhanced matching (Matching 3, $\delta=0.4$, $\delta_b=0.005$, $N_b=9$) on Figure \ref{figure:PerformanceThoughTime}.
The first general observation is that on the most part of the 5-year span of the database, the improvement due to the enhanced matching procedure is very significant.
Four global periods are easily identified. From Wednesday January 2nd, 2008 to Tuesday September 30th, 2009 (included), the basic matching can place roughly $45\%$ of market orders in the book, while the performance of the enhanced matching is roughly $70\%$\footnote{These performances might be improved by tuning the parameters, but this is not the aim of this section.}. Note that our database has a large gap of missing data in this period. On Wednesday October, 1st, 2009, the performance of the basic matching suddenly jumps to roughly $55\%$ and keeps this level until Friday June 12th, 2010 (included), while in this period the performance of the enhanced matching stays constant. On Monday June 14th, 2010, the performances of both matching algorithms jump to roughly $90\%$, and keep this level until Friday November 12th, 2010. During this five months, the performance of the basic algorithm is very close to the performance of the enhanced one. On Monday November 15th, 2010, however, the performance of the basic matching drops to roughly $85\%$ while our enhanced algorithm remains at roughly $90\%$. This observation remains valid until the end of the sample (June 2013).

It is very interesting to observe that all the three above dates are also ruptures for the stock ACCP.PA. They are therefore to be interpreted as exchange-related events.
On October 1st, 2009, Euronext made its "Universal Trading Platform - Market Data" its primary data feed for European markets\footnote{Euronext Info-flash dated September 30th, 2008}. We may conjecture that this change has been translated (through the data vendor building the database) into a significant improvement for the database.
On June 14th, 2010, Euronext significantly improved the way it disseminates the best bid and offer quotes through its Universal Trading Platform\footnote{Euronext Info-flash dated May 18th, 2010}, there again we may conjecture that this is the cause of a significant improvement of the quality of the historical tick data.
Finally, a precise cause for the observed drop for Matching 1 on November 15th, 2010 is less clear to conjecture, but more elements will be provided in Section \ref{section:TradeSignature}.

\begin{figure}
\begin{center}
\begin{tabular}{cc}
\includegraphics[scale=0.43, page=1]{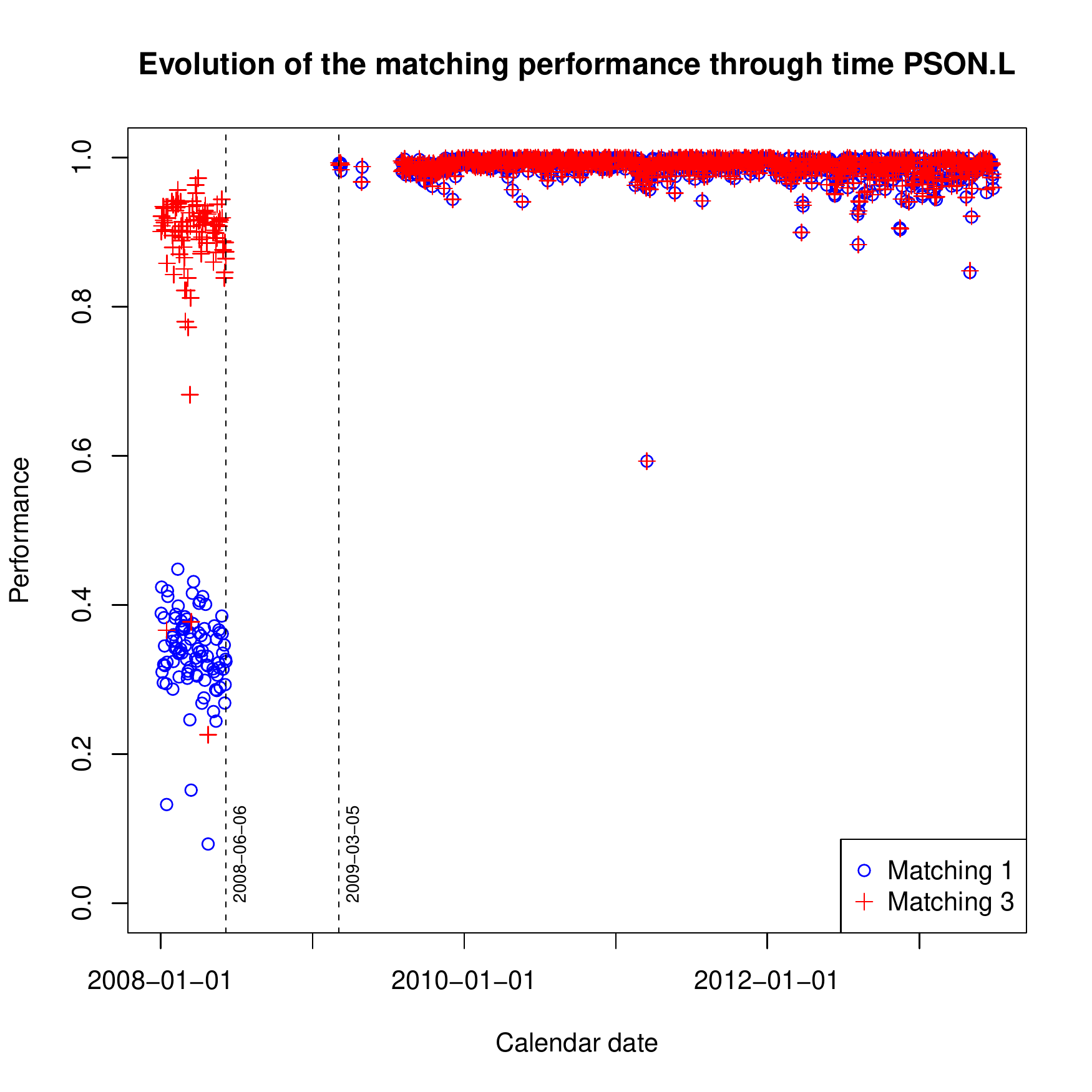}
&
\includegraphics[scale=0.43, page=1]{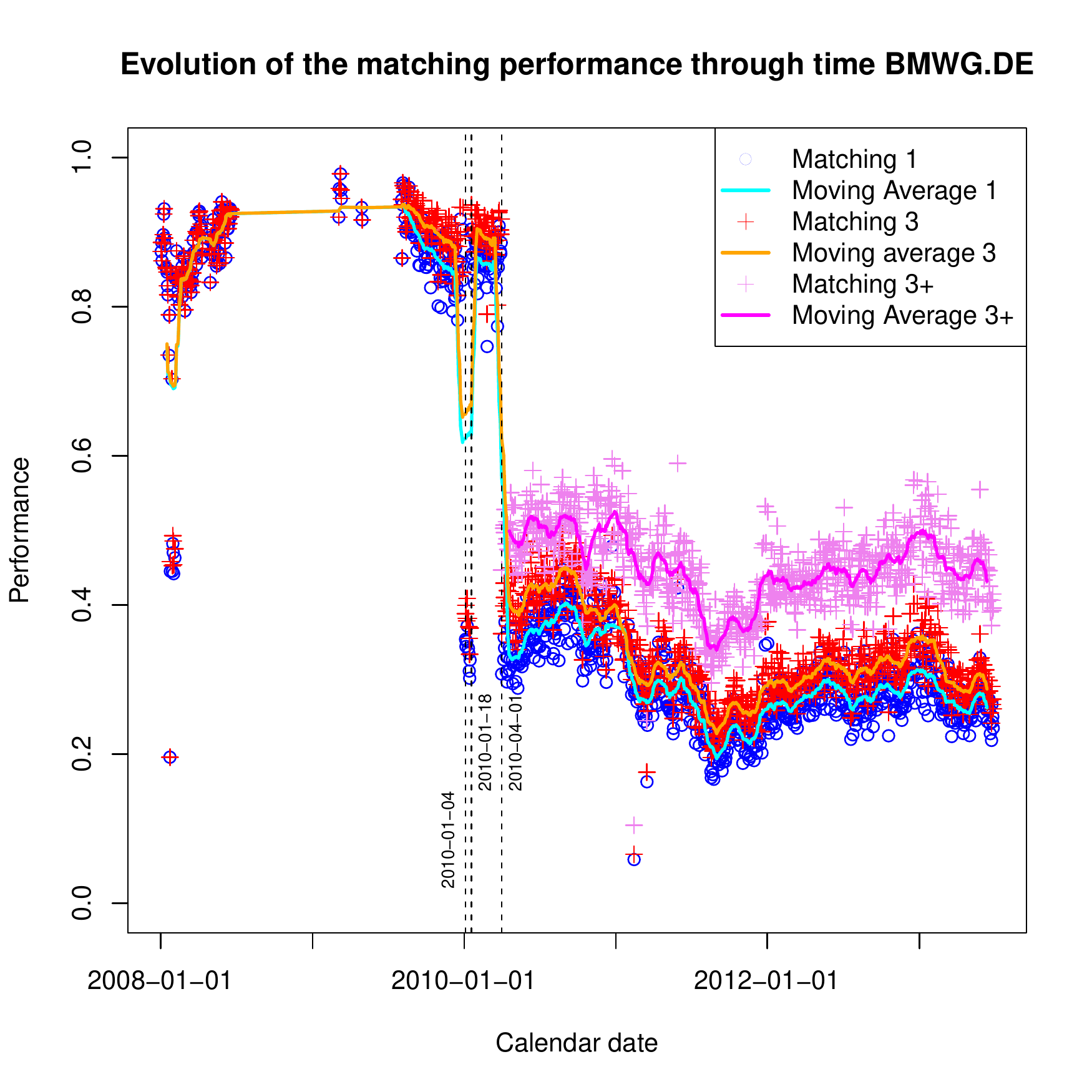}
\end{tabular}
\end{center}
\caption{Performance of the basic and enhanced matching algorithms for the stock PSON.L (left) and BMWG.DE (right) from Jan, 1st, 2008 to June, 30th, 2013, with gaps.
For PSON.L, $\delta=1.0$ in 2008, $0.4$ from 2009.}
\label{figure:PerformanceThoughTime-PSON-BMWG}
\end{figure}

On Figure \ref{figure:PerformanceThoughTime-PSON-BMWG}, we present a similar plot for two others exchanges, the London Stock Exchange (stock PSON.L, Pearson Group, component of the Footsie) and the Frankfurt Stock Exchange - Deutsche Börse (stock BMWG.DE, BMW AG, component of the DAX).
Several observations can be made. Firstly, these plots also display obvious ruptures in the matching process, i.e. in the quality of the database. For PSON.L, the basic matching with $\delta=0.4$ is nearly perfect for the major part of the database, starting March 2009. In 2008 however, this basic matching barely matches 1/3 of the orders even with larger parameters ($\delta=1$). This demonstrates the usefulness of our algorithm, even for exchanges it was not designed for, since our enhanced matching is able to match roughly $90-95\%$ of the trades even in 2008.
Due to a gap in our data, we are not able to exactly pinpoint the date of the quality change of the database, although we can conjecture, following the Euronext case above, that the "Market Data Optimisation 2009" project, that was designed in particular to improve data feeds and was precisely implemented during this period\footnote{London Stock Exchange Service Annoucements 125/08 and 130/08, December 2008.}, is a good candidate to explain this shift in performances.

As for the stock BMWG.DE, in opposition to the previous exchanges and very counter-intuitively, we observe a sharp degradation of the matching performances through time. Until April 1st, 2010, we observe matching performances that compare to the previous exchanges, using the same parameters used for the other stocks : both the basic and enhanced procedures are able to exactly match around $90\%$ of the transactions, the enhanced matching appearing slightly better starting from 2009. But on April 1st, 2010, the performance of the basic procedure suddenly drops to roughly $30\%$ of the trades. The enhanced procedure is still better, but even with bigger parameters (Matching 3+ uses $\delta_b=1.0$ second and $\delta=2.0$ seconds) its performance stays in the order of magnitude of $50\%$. The same drop had been observed during the first two weeks of 2010. 
It is interesting to note that April 1st, 2010 is precisely the date Deutsche Börse implemented significant changes in its data transmission, restricted the access to un-netted market data to its "Xetra Enhanced Broadcast Solution"\footnote{Xetra Circular 151/09 dated December 2nd, 2009.}. We can conjecture that these changes were the probable cause of the shift we observe on the quality of the database of our data vendor. Interestingly enough, the change was initially planned on January 4th\footnote{Xetra Circular 151/09 dated December 2nd, 2009.}, which reinforces our conjecture.

\subsection{Influence of the liquidity (number of transactions)}
\begin{figure}
\begin{center}
\includegraphics[scale=0.7, page=1]{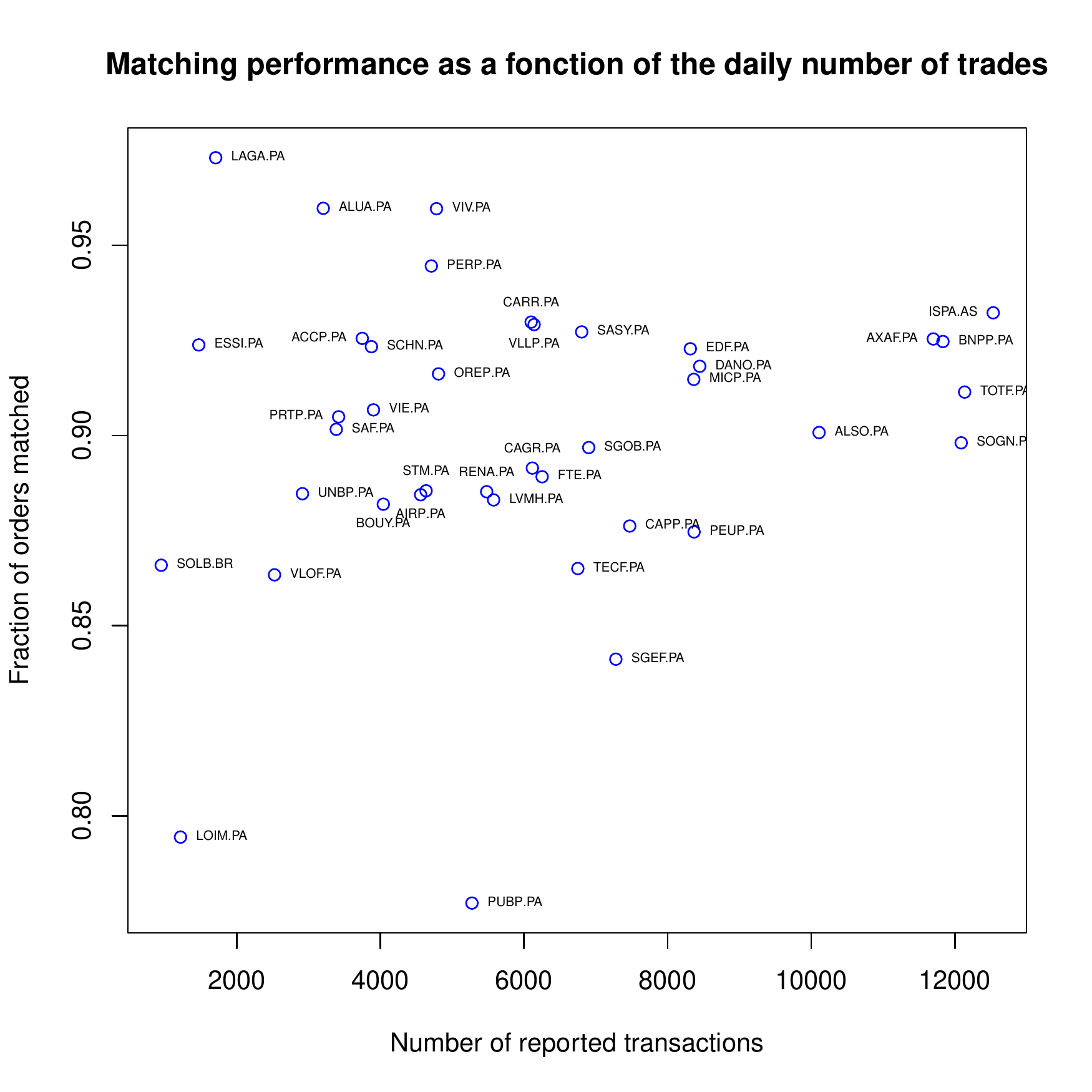}
\end{center}
\caption{Proportion of matched trades as a function of the total number of reported transactions. All trades reported in the database on Jan. 17th, 2011 between 09:05 and 17:20 are considered for matching. In this example, $\delta=0.4, N_b=9, \delta_b=0.005$.}
\label{figure:Liquidity}
\end{figure}

\begin{figure}
\begin{center}
\begin{tabular}{cc}
\includegraphics[scale=0.43, page=1]{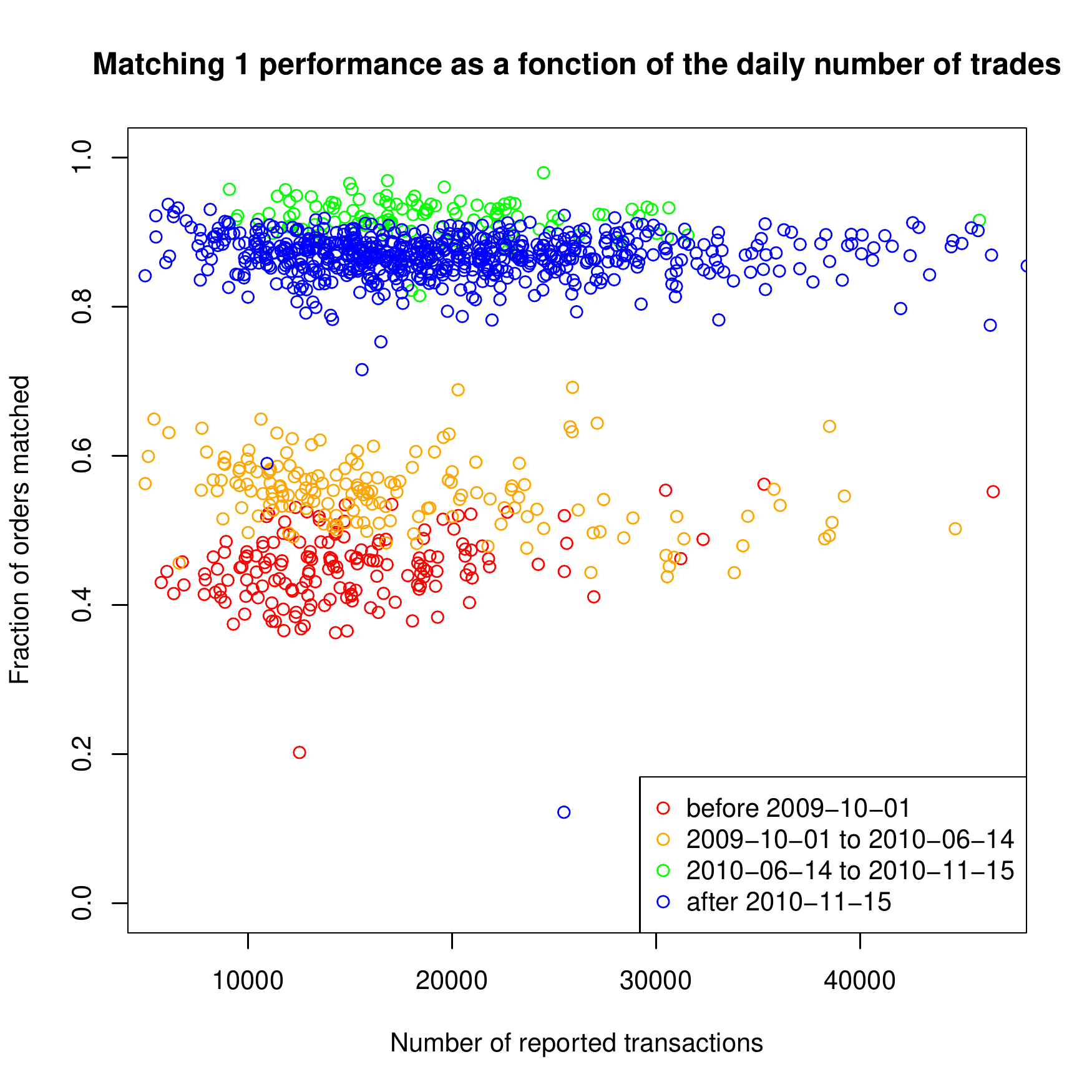}
&
\includegraphics[scale=0.43, page=2]{{Liquidity-BNPP.PA}.pdf}
\end{tabular}
\end{center}
\caption{Proportion of matched trades as a function of the total number of reported transactions for the stock BNPP.PA for both matching procedures for the several periods previously identified.}
\label{figure:Liquidity-BNPP.PA}
\end{figure}

We close this section by testing the naive intuition linking the quality of the matching to the number of trades on the studied stock: intuitively, for heavily traded stocks, overloads of the reporting system might occur, leading to potentially more inaccuracies in the "quotes" and "trades" files, so that the matching performance should decrease.
Figure \ref{figure:Liquidity} plots the matching performance as a function of the number of trades reported during one trading day for 40 different stocks mostly trade on Euronext, this sample being very close, but not identical, to the composition of the CAC 40 index. Names of the stocks are provided on the plot.
Complementary to this cross-stocks view, a time-dependent view of the same idea is given by Figure \ref{figure:Liquidity-BNPP.PA}, where the performance of the enhanced matching for all the trading days for the stock BNPP.PA are given as function of the number of trades reported in the "trades" file (and distinguishing between the four performance periods identified in Section \ref{subsection:MatchingPerformanceThoughTime}).
Both Figures \ref{figure:Liquidity} and \ref{figure:Liquidity-BNPP.PA} tend to visually discard the tested hypothesis: the scatter plots do not reveal any clear relationship between the performances and the number of trades. Therefore, at least on this sample, discrepancies between the "quotes" and "trades" files do not appear to be due to an overload of the reporting systems.

\section{Assessment of the performance of trade signature}
\label{section:TradeSignature}

\subsection{The Lee-Ready procedure}

Signing a trade consists in assigning to each recorded transaction a "buy" or "sell" label, depending on the side of the market order that initiated this transaction. A "buy" market order is matched to liquidity (pending limit orders) on the ask side of the order book, while a "sell" market order is matched to liquidity (pending limit orders) on the bid side of the order book.
One of the first notable algorithm for trade signing was proposed by \citet{LeeReady1991}. The motivation for this work at that time was to improve the standard signing techniques that roughly fell into two categories : the tick test, which decides whether a trade is buy or sell by comparison to adjacent trades (i.e. without using any quotes data) ; and the comparison to the current quote.
\citet{LeeReady1991} observe that the current quote is often not synchronized to the trades, so they suggest based on this empirical observation that if a quote is less that five seconds old, the trade price should be compared to the quote 5 seconds before.
This algorithm has encountered a certain success, being regularly used or revisited in the financial literature. Among others, \citet{Ellis2000} reports of roughly $80\%$ of trades correctly classified by the Lee-Ready algorithm using 1996-1997 NYSE data (the "true" sign is determined by using several data indicators, including the identity of the trader), \citet{theissen2001} reports a success rate of roughly $70\%$ using 1996 data on German DAX and MDAX stocks (the "true" sign is determined using market maker information). More recently, \citet{Chakrabarty2012} assess the performance of the Lee-Ready procedure by comparing its results to a sign provided by the exchange (i.e. not using quotes data) and find on 2005 data a misclassification by the Lee-Ready procedure of the order of magnitude of $20$ to $30\%$ (similar to \cite{omrane2016} on 2005 data as well). Please note these few references have been selected just as an illustration with various data (from 1996 to 2005), references in each of them should be checked for further details.

By parsing the "trades" file and matching it to the "quotes" file, our reconstruction of the order flow provides the signature of the trades as a by-product, so that we are in a position to contribute to the assessment of the Lee-Ready signing procedure.
Of course, the five seconds delay recommended in 1991 is probably not relevant anymore more than twenty years later in a context of high-frequency data, but the idea behind the procedure --- finding the optimal lag/quote for comparison --- is still valid (see e.g. \citet{Piwowar2006} using 1990s data or \citet{Chakrabarty2012} using 2005 data).
Therefore, we implement the Lee-Ready procedure as follows : for a trade with transaction time $\tau_t$, retrieve the best bid and ask quotes at time $\tau_q=\tau_q+\delta_{LR}$ ; if the trade price is strictly greater (resp. lower) than the mid-price, classify the trade as "buy" (resp. sell). If the trade is equal to the mid-price, we left its sign undecided. Of course, we could implement a tick test to decide this last case, and obtain the "full" Lee-Ready algorithm, but we coherently restrict ourselves to a quote-only procedure.

\subsection{Lee-Ready performance as a function of the lag}

\begin{figure}
\begin{center}
\begin{tabular}{cc}
\includegraphics[scale=0.4, page=1]{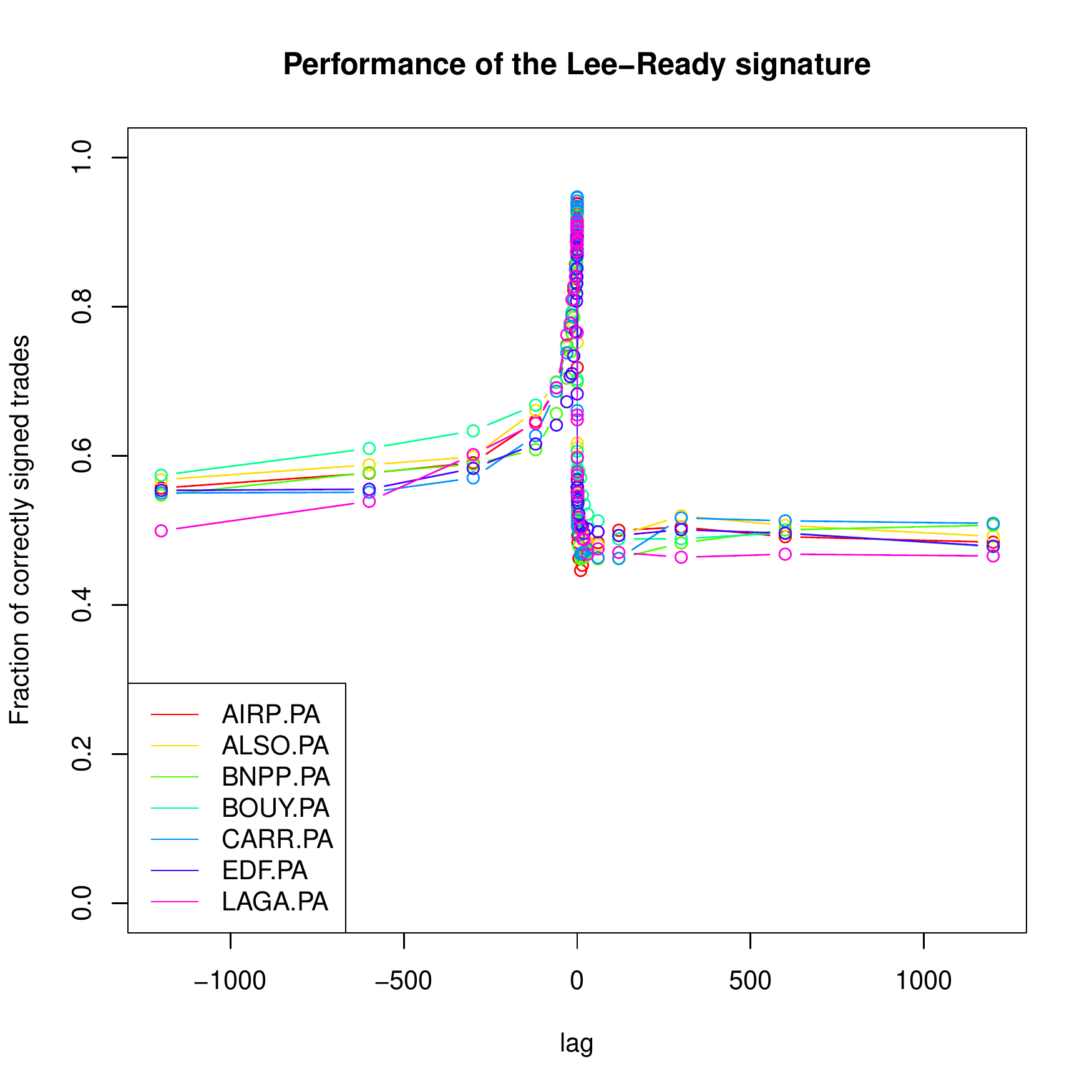}
&
\includegraphics[scale=0.4, page=2]{LeeReady}
\\
\multicolumn{2}{c}{\includegraphics[scale=0.4, page=3]{LeeReady}}
\end{tabular}
\end{center}
\caption{Performance of the Lee-Ready signing procedure (fraction of correctly signed orders) as a function of the lag parameter $\delta_{LR}$, in seconds. All three plots represent the same data, at a different zooming level around $0$. }
\label{figure:LeeReady}
\end{figure}

We now compare the results of the Lee-Ready procedure to the result of our own matching procedure. Our signing is considered the "true" signing, since it fully uses the "quotes" data, including available liquidity volumes. Orders that are not signed by our procedure are excluded from the comparison.
Figure \ref{figure:LeeReady} plots the fraction of the correctly signed orders by the Lee-Ready procedure as a function of the lag parameter $\delta_{LR}$ (ranging from $-10$ minutes to $+10$ minutes).

The first observation is that the global performance of the Lee-Ready signature is quite good, even better than most reports on this procedure. On our 7-stock 1-day sample used above in Section \ref{section:ReconstructionOrderFlow}, the Lee-Ready procedure correctly signs roughly $90\%$ of the trades when using the current quotes or a lagged quote with a small lag (a few tens or hundreds of milliseconds).
When using a larger negative lag for the signing quote, the performance drops quite fast. Between $-5$ and $-10$ seconds lag, the signature performance is roughly $80\%$ on our sample.
The $90\%$ performance obtained in this sample is slightly better than what is usually reported in the financial literature, but note that most previous works use older data and do not go below a granularity of $1$ second. It is quite intuitive that recent tick-by-tick data timestamped to the millisecond should allow for a better signature performance.

The second observation is that, at least visually on this small sample, the lag rule does not seem to be increasing the performance. The five second rule of 1991 is not relevant on 2011 high-frequency data, but even very small lags slightly depreciate the performance. The third panel of Figure \ref{figure:LeeReady} shows the slight performance drop as the lag goes from $0$ to $-200$ milliseconds.
This would tend to advocate for the use of contemporaneous quotes when signing trades with the Lee-Ready algorithm on \emph{recent} data, but we will provide a more detailed view on this problem in Section \ref{subsection:SignaturePerformanceThtoughTime}.

Our third observation is that this signing performance drops dramatically for $\delta_{LR}\in[0.050, 0.150]$ before eventually increasing again for larger positive lags. In roughly $100$ milliseconds, the signature performance drops from roughly $90\%$ to roughly $60\%$. This allows us to make a clear link to Figure \ref{figure:TimeLagMatching1} : the time interval $[0.050, 0.150]$ is precisely the support of the synchronization lag between the "trades" file and the "quotes" file for our 7-stock 1-day sample under scrutiny here. Therefore during that time interval the transactions are reflected on the quotes, and the prices changes occur, if any, so that the quotes after that reflect the consequences of the trades, and therefore cannot be used to sign them.
In other words, Figure \ref{figure:LeeReady} is an illustration of the market impact of trades (see e.g. \cite{Bouchaud2009} for a review). It is particularly interesting to see that the performance of the Lee-Ready signing procedure actually reflects the instantaneous market impact of the trades: immediately after the transaction, prices may move --- so that the quotes are not useful to sign the trades --- but the impact gradually disappear --- the performance of the signature increases slowly for large positive $\delta_{LR}$.

\subsection{A toy model for the signature performance}

We now propose a toy model that account for the shape of the performance of the Lee-Ready signature as observed on Figure \ref{figure:LeeReady}. 
Let $m_t$ be the mid-price at time $t$.
Let us assume that the flows of orders that moves the mid-price are homogeneous Poisson processes : let $\lambda^{LC}_+$ (resp. $\lambda^{M}_+$) be the intensity of the limit orders and cancellations (resp. market orders) that make the mid-price move up, and $\lambda^{LC}_-$ and $\lambda^{M}_-$ be the intensities for the same events that instantaneously trigger downward moves of the mid-price.
For simplicity, we assume that all movements of the best quotes have a size of one tick, and that the spread is constantly equal to one tick, so that the mid-price also moves by exactly one tick.

Let us now take a market order that is incorporated in the "quotes" file at time $\tau_q$. This market order is (has been most of the times) reported in the "trades" files at time $\tau_t=\tau_q-\delta$ where $\delta$ is for now assumed to be deterministic.
Furthermore, when attempting to sign this trade, the Lee-Ready algorithm will compare it to the mid-price at time $\tau_{LR}=\tau_t+\delta_{LR}=\tau_q+(\delta_{LR}-\delta)$.

Assume for now that $\tau_{LR} < \tau_{q}$, i.e. $\delta-\delta_{LR}>0$, and that we have a buy market order. It is straightforward to see that this buy market order will be correctly signed by the Lee-Ready signature if and only if $m_{\tau_{LR}} \leq m_{\tau_q}$ i.e. if and only if $N_+\geq N_-$ where $N_+$ is the number of events on the time interval $(\tau_{LR},\tau_{q})$ making the mid-price move upward and $N_-$ is the number of events making the mid-price move downward on this same time interval. Since $N_+$ and $N_-$ are Poisson variables, their difference is distributed according to a Skellam (or Poisson Difference) distribution.
Recall that the cumulative distribution function of the Skellam distribution with parameters $(\mu_1,\mu_2)$ is written :
\begin{equation}
	F(n ; \mu_1,\mu_2) = e^{-(\mu_1+\mu_2)} \sum_{k=-\infty}^n \left(\frac{\mu_1}{\mu_2}\right)^{k/2}I_{k}(2\sqrt{\mu_1\mu_2}),
\end{equation}
where $I_{k}$ is the modified Bessel function of the first kind.
Then the probability $p^b(\delta, \delta_{LR})$ that the buy market order considered is correctly signed by the anterior $\tau_{LR}$ mid-price is :
\begin{equation}
	p_<^b(\delta, \delta_{LR}) = F(0 ;
		(\lambda^{LC}_- +\lambda^{M}_-)|\delta-\delta_{LR}|,
		(\lambda^{LC}_+ +\lambda^{M}_+)|\delta-\delta_{LR}| ).
\end{equation}
In the case of a sell market order, we obviously need $N_+-N_-\leq 0$, hence we obtain the result by exchanging the $+$ and $-$ intensities:
\begin{equation}
	p_<^s(\delta, \delta_{LR}) = F(0 ;
		(\lambda^{LC}_+ +\lambda^{M}_+)|\delta-\delta_{LR}|,
		(\lambda^{LC}_- +\lambda^{M}_-)|\delta-\delta_{LR}| ).
\end{equation}

Let us now turn to the case $\tau_{q} < \tau_{LR}$, i.e. $\delta-\delta_{LR}<0$. Let us first assume that the market order we consider does \emph{not} change the mid-price. In this case, with the same arguments as above, a buy (resp. sell) market order will be correctly signed if and only if $m_{\tau_q} \geq m_{\tau_{LR}}$ (resp. $m_{\tau_q} \leq m_{\tau_{LR}}$) i.e. if and only if $N_+ - N_- \leq 0 $ (resp. $N_- - N_+ \leq 0 $) on the time interval $(\tau_{q},\tau_{LR})$.
In short, we have with obvious notations the probability to correctly sign non quote-changing trades with posterior quotes :
\begin{align}
	p_>^b(\delta, \delta_{LR}) & = p_<^s(\delta, \delta_{LR}),
	\\
	p_>^s(\delta, \delta_{LR}) & = p_<^b(\delta, \delta_{LR}).
\end{align}

Now, if we consider a trade that changes the best quote (by assumption by exactly one tick), we need to correct for its instantaneous impact. Therefore, with the same arguments as above, a buy (resp. sell) market order will be correctly signed if and only if $m_{\tau_q} \geq m_{\tau_{LR}}$ (resp. $m_{\tau_q} \leq m_{\tau_{LR}}$) i.e. if and only if $(N_+ + 1)- N_- \leq 0 $ (resp. $(N_- +1) - N_+ \leq 0 $) on the time interval $(\tau_{q},\tau_{LR})$.
Therefore, in this case :
\begin{align}
	p_{>,agg}^b(\delta, \delta_{LR}) & = F( -1 ;
		(\lambda^{LC}_+ +\lambda^{M}_+)|\delta-\delta_{LR}|,
		(\lambda^{LC}_- +\lambda^{M}_-)|\delta-\delta_{LR}| ),
	\\
	p_{>,agg}^s(\delta, \delta_{LR}) & = F( -1 ;
		(\lambda^{LC}_- +\lambda^{M}_-)|\delta-\delta_{LR}|,
		(\lambda^{LC}_+ +\lambda^{M}_+)|\delta-\delta_{LR}| ).
\end{align}

Finally, let $\rho_{agg}$ be the fraction of (aggressive) market orders (of either side) that make the price move (equivalently let us assume that each market order can make the mid-price move with probability $\rho_{agg}$). By basic properties of Poisson processes the fraction of buy market orders is $\rho_+ = \frac{\lambda^M_+}{\lambda^M_+ +\lambda^M_-}$ since among market orders only buy (resp. sell) orders can make the price move up (resp. down).
Then the probability $p(\delta,\delta_{LR})$ that a trade, reported in the quotes file with a lag $\delta$ from the "trades" file timestamp, is correctly signed by the "quotes" taken with a lag $\delta_{LR}$ from the timestamp of the "trades" file, is computed as :
\begin{align}
	p(\delta,\delta_{LR}) =
	& \mathbf 1(\delta-\delta_{LR}>0)
		\left[ 
			\rho_+ p_<^b(\delta, \delta_{LR})
			+ (1-\rho_+) p_<^s(\delta, \delta_{LR})
		\right]
	\nonumber \\
	&	+ \mathbf 1(\delta-\delta_{LR}<0)
		\left[
			(1-\rho_{agg})\left[ 
				\rho_+ p_>^b(\delta, \delta_{LR})
				+ (1-\rho_+) p_>^s(\delta, \delta_{LR})	
			\right]
		\right.
	\nonumber \\
	&	\left.
			+ \rho_{agg} \left[ 
				\rho_+ p_{>,agg}^b(\delta, \delta_{LR})
				+ (1-\rho_+) p_{>,agg}^s(\delta, \delta_{LR})	
			\right]
		\right] .
\label{eq:deterministicPerf}
\end{align}
Finally, we can now incorporate in the model the fact that $\delta$ is a random variable with an probability density function that should resemble the empirical densities of Figure \ref{figure:TimeLagMatching1}. If $\delta$ is assumed to be absolutely continuous with density $f_{\delta}$, then the theoretical performance of the Lee-Ready signature using a $\delta_{LR}$ lagged-quote is:
\begin{equation}
	p(\delta_{LR}) = \int p(u,\delta_{LR}) f_{\delta}(u) du. 
\label{eq:randomPerf}
\end{equation}

\begin{figure}
\begin{center}
\begin{tabular}{cc}
\includegraphics[scale=0.43, page=1]{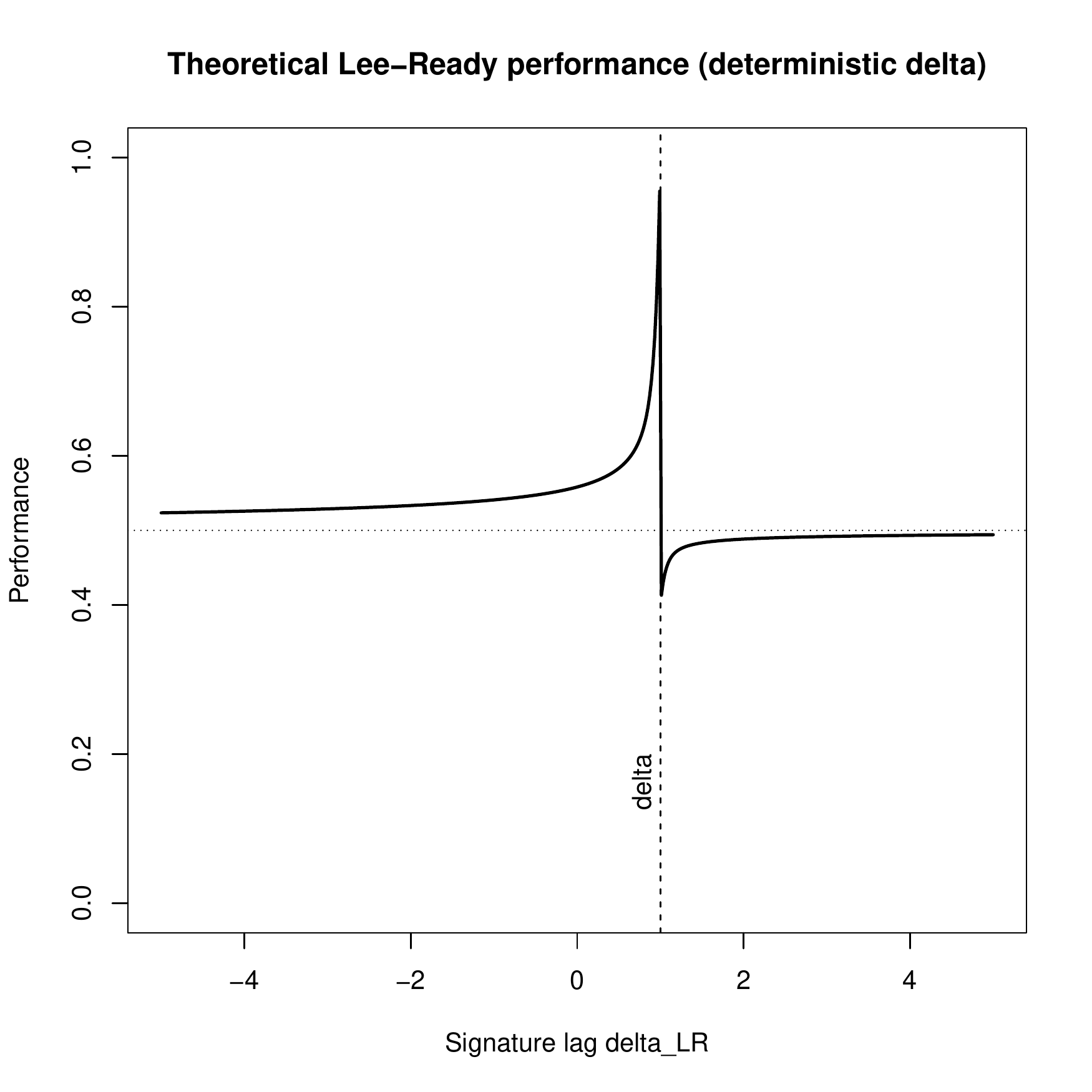}
&
\includegraphics[scale=0.43, page=2]{LeeReady-Skellam-Spread1}
\end{tabular}
\end{center}
\caption{Theoretical performance of the Lee-Ready signature in the proposed toy model, for a deterministic $\delta$ (left) and a random $\delta$ (right). For these illustrations,$\lambda^{LC}_\pm=5$, $\lambda^M_\pm=1$, $\rho_{agg}=0.6$ and $\delta=1$ (deterministic) or $\delta$ is Gaussian with parameters $(1;0.1)$ (random).}
\label{figure:LeeReady-Skellam}
\end{figure}

\begin{figure}
\begin{center}
\begin{tabular}{cc}
\includegraphics[scale=0.43, page=4]{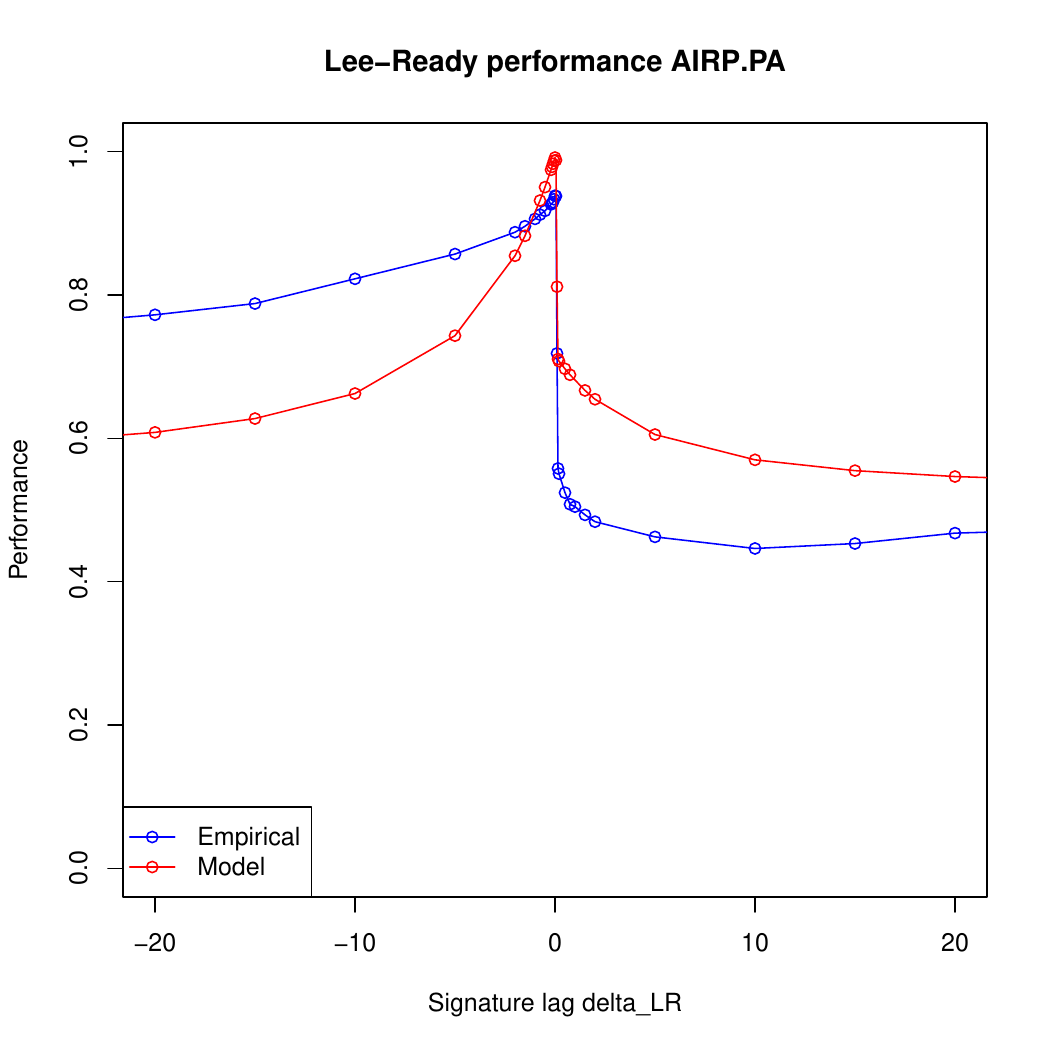}
&
\includegraphics[scale=0.43, page=5]{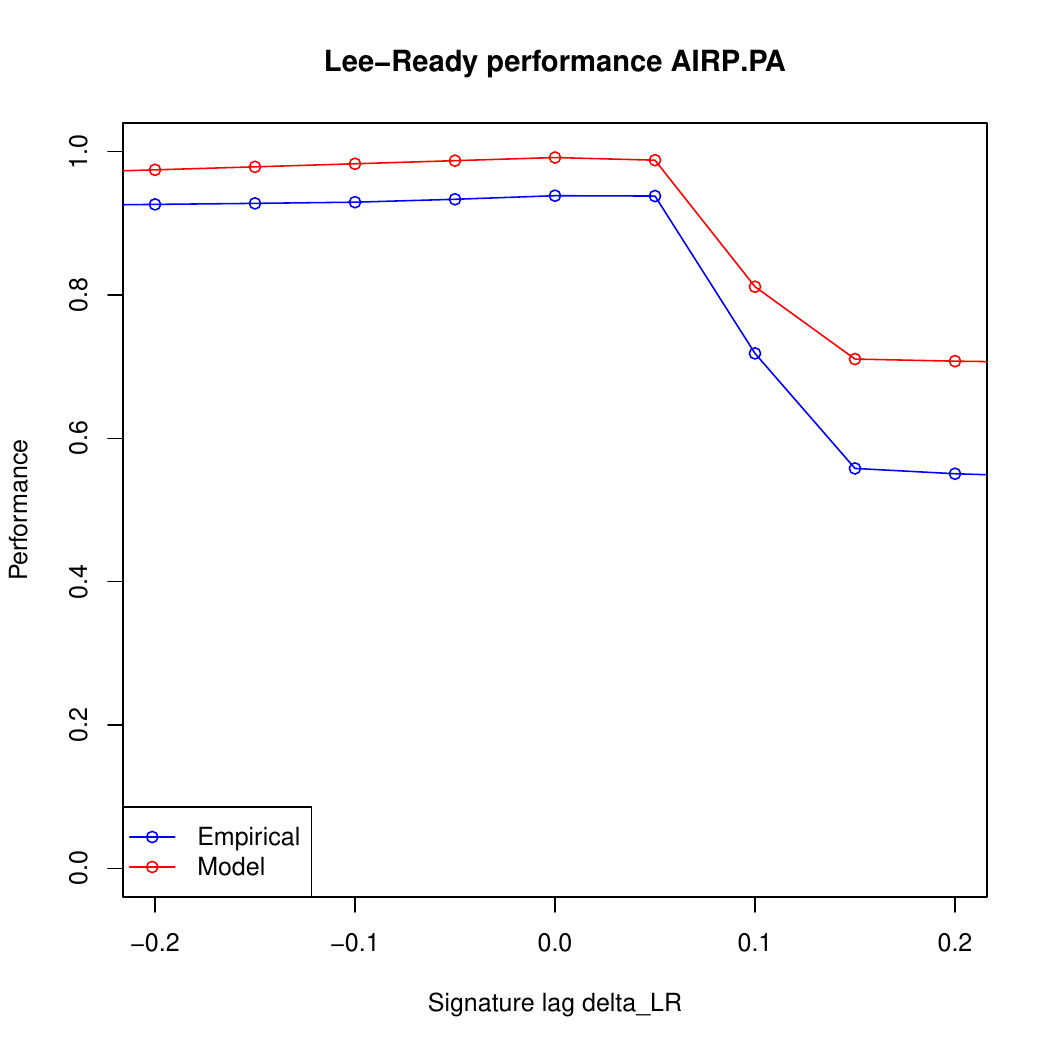}
\end{tabular}
\end{center}
\caption{Empirical vs toy model performance of the Lee-Ready signature for the stock AIRP.PA. Both left and right graphs show the same data at different time scales.}
\label{figure:LeeReady-Skellam-Calibration}
\end{figure}

Figure \ref{figure:LeeReady-Skellam} plots examples of these theoretical performances, illustrating how Equation \eqref{eq:randomPerf} is able to account for the general shape of the Lee-Ready performance as a function of the lag, as it is observed on our sample.
As this is a toy model however, based on time-homogeneous Poisson processes that are not able to capture the features of the order flows (autocorrelation of trades signs, trades clustering, etc.), we should not expect high fitting power when calibrated on market data. The results for the fitting on our 7-stock 1-day sample gives nonetheless interesting results.
Figure \ref{figure:LeeReady-Skellam-Calibration} plots the results for the stock AIRP.PA (for the sake of brevity we reproduce only the graphs for the first stock of the sample, but they are all similar).
Firstly, at small time scales, the model is able to satisfactorily reproduce the drop of the signature performance around $\delta_{LR}=+100$ milliseconds, translating the incorporation of the trades into the quotes. 
Secondly, at larger time scales, we observe that the empirical performance is globally overestimated by the model from roughly $\delta_{LR}=-1$ second and onwards. This might be a mark of the clustering of market events: trades are submitted irregularly during the trading day, with burst of activities during which all types of market events occur ; it is therefore expected that the mid-price moves more actively around the timestamps of trades submission, decreasing the actual empirical signature performance. One may also argue that the fact that the overestimation of the performance by the model is greater after the trades than before is an indication that transactions trigger market activity rather than result from it.

\begin{remark}
It is easy to make the model more flexible by adding a spread distribution or by distinguishing between aggressive buy and sell orders. However, the roughness of the Poisson hypothesis is such that these enhancements are not expected to be crucial.
\end{remark}

\subsection{Signature performance through time}
\label{subsection:SignaturePerformanceThtoughTime}

\begin{figure}
\begin{center}
\begin{tabular}{cc}
\includegraphics[scale=0.43, page=1]{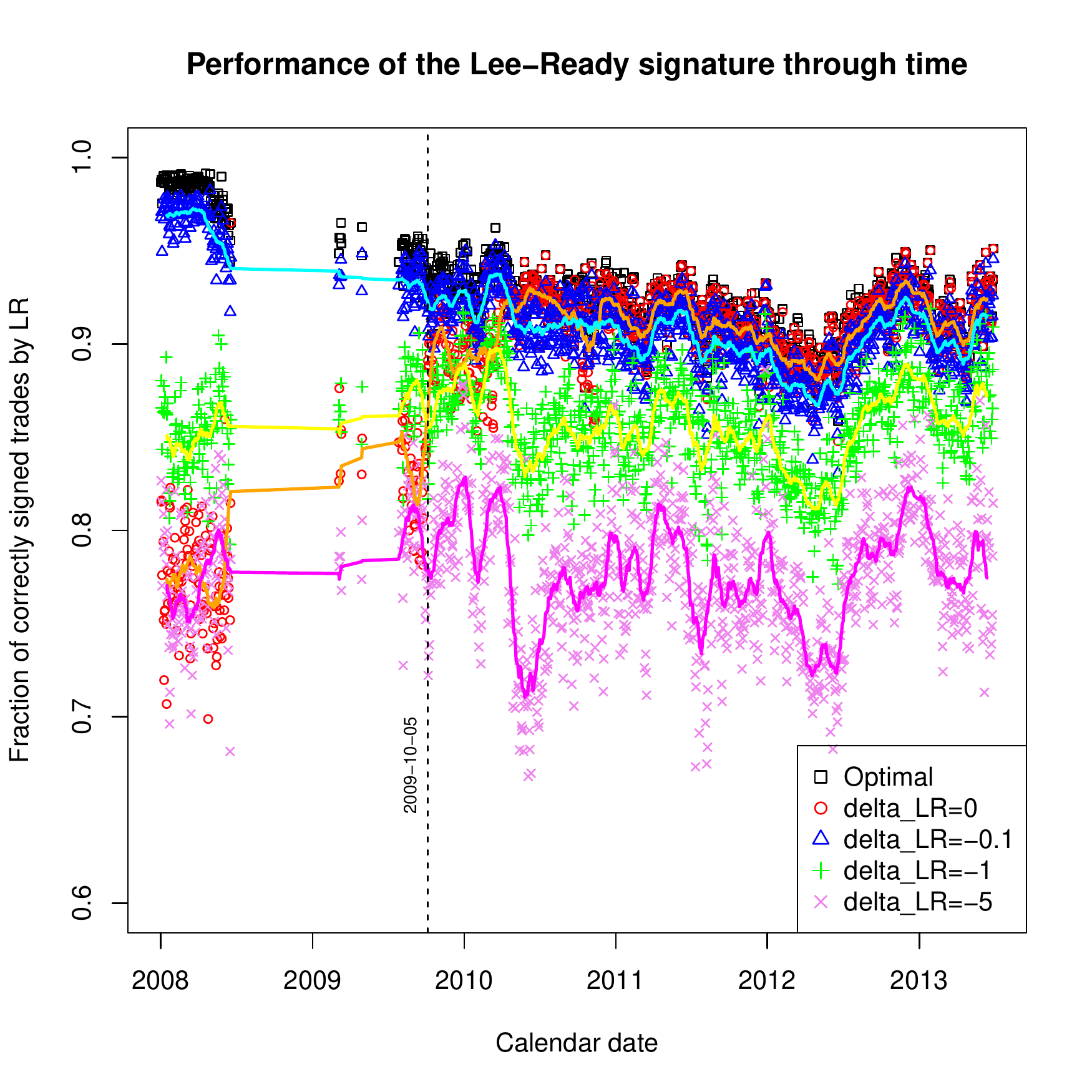}
&
\includegraphics[scale=0.43, page=3]{LeeReadyThroughTime}
\end{tabular}
\end{center}
\caption{Performance of the Lee-Ready signature for the stock BNPP.PA (left) and optimal lag for these signatures (right) as a function of the calendar date. Continuous lines are monthly moving average of the associated quantities ($\pm 11$ trading days).}
\label{figure:LeeReadyThroughTime}
\end{figure}

To complement the previous results, we run the Lee-Ready procedure on the stock BNPP.PA for all the trading days of the database, with lags $\delta_{LR}\in\{-5, -2, -1.5, -1, -0.75, -0.5, -0.2,$ $-0.15, -0.1, -0.05, 0, 0.05, 0.1\}$ seconds. On Figure \ref{figure:LeeReadyThroughTime} we plot the (\emph{a posteriori}) optimal performance of the Lee-Ready signature as a function of the calendar dates, as well as the performance for the representative lags $-5,0-1,-0.1$, and $0$. On the right pannel, we plot the optimal lag (giving the optimal performance) for each of these days.
The first observation is that signature performance by the Lee-Ready algorithm are roughly constant over time, with (visually) a slight downward trend of the optimal performance from 2008 to 2012 and a increasing slope for the last part of the sample, but roughly always staying around $90\%$ and above at all time
The second observation is that performances for given lags do not move in parallel. In 2008, the use of the current quote ($\delta_{LR}=0$) and the use of a $-5$ seconds lagged quote would give similar performances, less that $80\%$, while the use of the $-1$ second lagged quote would improve the result (in accordance with \citet{Chakrabarty2012} on 2005 data), and the use of the $100$ milliseconds lagged quote would even be better.
One therefore cannot recommend absolute values for the Lee-Ready lag, since these quantities depend intrinsically on the database used, the exchange, the stock and the time-window considered. From Figure \ref{figure:LeeReadyThroughTime}, we can only confirm the conjecture that the more recent the data studied, the shorter the lag should be used. On our sample, from 2011 and onwards, we even observe that a positive $\delta_{LR}=+50$ milliseconds is most of the times optimal to get the best performance of the Lee-Ready signature !

Very interestingly, the graphs allow as well for some forensic analysis of the quality of the data, as it is plausible that these variations in signature performances are marks of technical changes in the production of the data.
We observe that the performance of the signature with the current quotes ($\delta_{LR}=0$) sharply and suddenly increases on October 5th, 2009, a date close to one we have already identified in Section \ref{section:EmpiricalResults} as a change in data feeds on Euronext.
Furthermore, this analysis confirms that 2010 has been a year of significant changes in the production of the data, especially between April 21st, 2010 (the first day of the sample for which $\delta_{LR}=0$ gives the optimal signature performance, if we except one outlier in 2008) and November 15th, 2010 (the first day of the sample for which $\delta_{LR}=+0.050$ gives the optimal signature performance).
These dates are also very close to dates identified in Section \ref{section:EmpiricalResults}, and interestingly, they are also the dates of the migration process conducted by Euronext to relocate its data centre to Basildon (near London). This process started in April, with the first tests in May, and the production migration was completed in October\footnote{Euronext Info-flashes date April 21st, 2010, May 17th, 2010 and October 11th, 2010.}. We may conjecture that such a huge technical change is the source of our observations, either directly or through changes that the data vendor (Reuters) would implement in response.

It is striking that a simple operation such as trade signature bears the marks of these technical changes.

\section{Potential side effects of choices of order flow reconstruction}
\label{section:SideEffects}

We end this work by an illustration of the potential side effects of choices made when reconstructing the order flow. Let us imagine the (very academic and hypothetical...) case of a trader looking to calibrate on market data a Hawkes process to model the clustering behaviour of trades. Hawkes models have been increasingly popular in finance and a vast literature is available on the subject (see e.g. \cite{Bowsher2007,Large2007} for pioneering papers and \cite{Bacry2015} for a recent review). Let us assume that our trader wants to model the counting process of market orders $(N^M_t)_{t\geq 0}$ with an intensity $\lambda^M_t$ driven by the following equation :
\begin{equation}
	\lambda^M_t = \lambda_0 + \int_0^t \alpha e^{-\beta(t-s)} dN^M_s,
	\label{equation:HawkesMarket}
\end{equation}
i.e. to use a one-dimensional Hawkes model with a unique exponential kernel. In this model, each market order makes the intensity increase with a jump of size $\alpha$, and this effect decreases exponentially fast with time with parameter $\beta$. The ratio $\frac{\alpha}{\beta}$ is interpreted as the fraction of trades triggered by previous trades in this model, in other words as a level of "endogeneity" in the market \citep{Filimonov2012,Hardiman2013}.
We are aware that this is arguably a very poor model to try on a full trading day --- a single exponential kernel, i.e. a single time scale, combined with a constant baseline intensity cannot satisfyingly model the complexity of the market, e.g. account for daily seasonality, see \citet{Challet2016} ---, but we will nonetheless keep this simple example for the sake of illustration. Our trader has two choices : the first one is to take the "trades" file as is, each different time stamp counting for one market order ; the second one is to precisely match each line in the "quotes" file, and in this case several lines might possibly be merged into one quote change, i.e. one transaction. In both cases, the model can then be fitted by numerical maximization of the log-likelihood of the model (see e.g. \cite{MuniToke2012} and references therein).

\begin{figure}
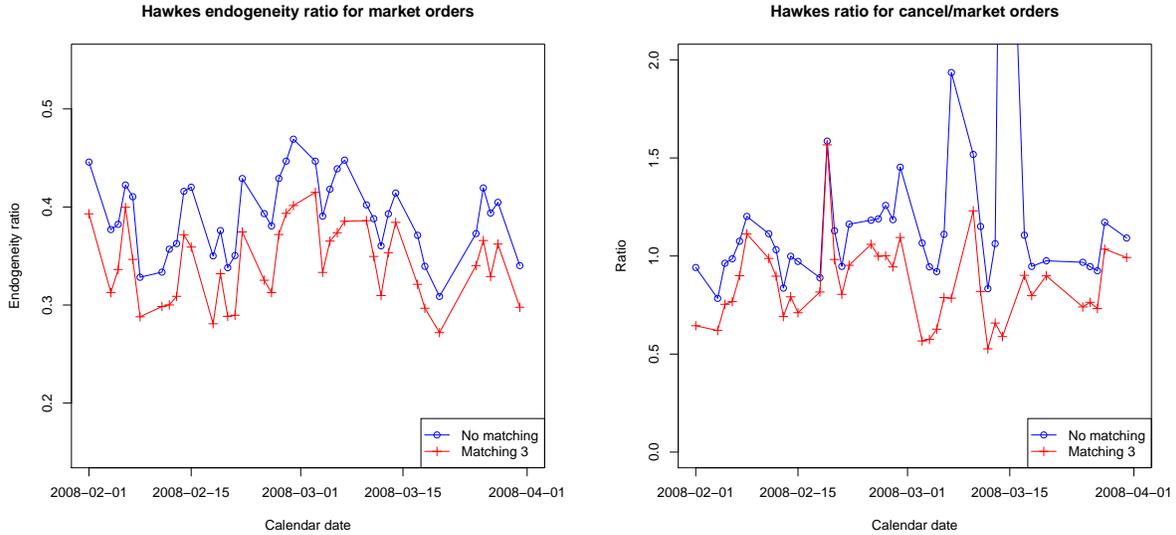

\begin{center}
\begin{tabular}{cc}
\includegraphics[scale=0.43, page=3]{{PSON.L-HawkesResults}.pdf}
&
\includegraphics[scale=0.43, page=4]{{PSON.L-HawkesResults}.pdf}
\end{tabular}
\end{center}
\caption{Hawkes ratio $\frac{\alpha}{\beta}$ calibrated for the model \eqref{equation:HawkesMarket} on the stock PSON.L in January and February 2008 (left). Idem for model \eqref{equation:HawkesCancelBQMarket} (right).}
\label{figure:PSON.L-HawkesResults}
\end{figure}

On Figure \ref{figure:PSON.L-HawkesResults}, we plot the ratio evolution $\frac{\alpha}{\beta}$ through time for the stock PSON.L in February and March 2008, i.e. two full trading months.
The plot shows that the endogeneity ratio estimated with the raw "trades" file is consistently greater by roughly $15\%$ to the endogeneity ratio estimated on a sample prepared with the matching procedure. Indeed, this result is consistent with the fact that matching procedure groups trades with close time stamps into one market order, if this is necessary to match the "quotes" file, while the calibration on the raw "trades" file considers these parts of the same trade as several closely-occurring market orders, therefore increasing the apparent clustering of transactions, hence the larger endogeneity ratio. This effect is best seen on the sample chosen since on this stock and at these dates, Matching 3 procedure groups a significant proportion of lines of the "trades" files, as is explicit on Figure \ref{figure:PerformanceThoughTime-PSON-BMWG}.

The matching procedure also has an effect on the cancellations, since transactions are matched against (a priori) cancellations at the best quotes. Let us now imagine that our trader were now to model the intensity $\lambda^C$ of the counting process of cancellations at the best quotes being excited by arrival of market orders (counting process $N^M$) as:
\begin{equation}
	\lambda^C_t = \lambda_0 + \int_0^t \alpha e^{-\beta(t-s)} dN^M_s.
	\label{equation:HawkesCancelBQMarket}
\end{equation}
There again, this arguably very poor model is kept simple for the purpose of illustration. The fitted ratio $\frac{\alpha}{\beta}$ in this case is plotted on Figure \ref{figure:PSON.L-HawkesResults}. In this case, the sample that does not match the "trades" and "quotes" overestimates the number of cancellations at the best quotes by keeping incorrect cancellations following trades that should be identified as the incorporation of the trades effect in the quotes. There is thus an incorrectly high number of occurrences of the pattern "trade then cancellation", it is therefore not surprising to observe here again a greater ratio when fitting without matching, that when fitting with prior matching.

\section{Conclusion}
We have investigated in this paper the TRTH trades and quotes database. We provide an algorithm that helps synchronizing the two data sources and therefore allows for a "true" reconstruction of the flow of limit orders, market orders and cancellations, even if we have only aggregated data and not any detailed information on the trades. It turns out that the enhanced matching is very useful for older data (2008), and still significantly better even for recent data (2013). We have provided basic examples illustrating potential side effects of the reconstruction procedure on model calibration. This hopefully may be useful for the community of researchers that work on high-frequency data but do not have access to detailed data down to the granularity of the messages sent to/from the exchange. 

This work is also a tentative of forensic analysis of financial data in the sense that we are able to track through our tick-by-tick aggregated data fundamental technical changes that occurred between 2009 and 2011 on the way financial exchanges (Paris, London, Frankfurt are tested here) produce data. It is very interesting to observe that these technical changes still leave marks on our aggregated database, even after several steps of data transformation and formatting.

As a conclusion, this work may be useful in reminding that a database, even packaged and provided by a unique source, is not a uniform object. Therefore, despite the temptation to use always larger sets of data in a search of statistical significance, the coherence of the sample should be ensured foremost.

\section*{Acknowledgements}
The author wishes to thank Pierre Garreau (Liquid Labs, Germany), Frédéric Abergel and Kévin Primicerio (CentraleSupélec, France) for stimulating discussions and comments.

\bibliographystyle{agsm}{}
\bibliography{OrdersFlowsUsingAggregatedData}

\end{document}